\documentclass[useAMS,usenatbib]{mn2e}

\usepackage{graphicx}

\title[Modelling the outflow in NGC\,1068]{Modeling the
[Fe\,II]$\lambda$1.644$\mu$m outflow and comparison with H$_2$ and
H$^+$ kinematics in the inner 200\,pc of NGC\,1068}
\author[F. K. B. Barbosa et al.]{
F. K. B. Barbosa$^{1}$\thanks{E-mail: fausto.barbosa@restinga.ifrs.edu.br},
T. Storchi-Bergmann$^{2}$,
P. McGregor$^{3}$ and
T. B. Vale$^{4}$
\\
$^{1}$IFRS -- Campus Restinga,
Rua 7121, n. 285 Lote 16, Quadra F, Restinga, CEP 91791-508, Porto Alegre, RS, Brazil
\\
$^{2}$Instituto de F\'{i}sica -- UFRGS, Caixa Postal 15051, CEP
91501-970, Porto Alegre, RS, Brazil
\\
$^{3}$Research School of Astronomy and Astrophysics, Australian National
University, Cotter Road, Weston Creek, ACT 2611, Australia
\\
$^{4}$INFES -- UFF, Rua Jo\~ao Jasbick, s/n, Bairro Aeroporto, CEP 28470-000
Santo Ant\^onio de P\'adua, RJ, Brazil
}
\begin{document}

\date{Accepted 2012 April 15. Received 2012 February 14; in original
form 2012 February 11}

\pagerange{\pageref{firstpage}--\pageref{lastpage}} \pubyear{2013}

\maketitle

\label{firstpage}

\begin{abstract}

We map the kinematics of the inner (200\,pc) narrow-line region
(NLR) of the Seyfert\,2 galaxy NGC\,1068 using the instrument NIFS
and adaptative optics at the Gemini North Telescope.
Channel maps and position-velocity diagrams are presented at a
spatial resolution of $\cong$\,10\,pc and spectral resolution $\sim
5300$ in the emission lines [Fe\,{\sc ii}]\,$\lambda\,1.644\,\mu$m,
H$_2\,\lambda\,2.122\,\mu$m and Br$\gamma$.
The [Fe\,{\sc ii}] emission line provides a better coverage of the
NLR outflow than the previously used [O\,{\sc iii}]\,$\lambda$\,5007
emission line, extending beyond the area of the bi-polar cone
observed in Br$\gamma$ and [O\,{\sc iii}].
This is mainly due to the contribution of the redshifted channels to
the NE of the nucleus, supporting its origin in a partial ionized
zone with additional contribution from shocks of the outflowing gas
with the galactic disc.
We modeled the kinematics and geometry of the [Fe\,{\sc ii}]
emitting gas finding good agreement with the data for outflow models
with conical and lemniscate (or hourglass) geometry.
We calculate a mass outflow rate of
$1.9^{+1.9}_{-0.7}$\,M$_\odot$\,yr$^{-1}$ but a power for the
outflow of only $0.08\%$~L$_{Bol}$.
The molecular (H$_2$) gas kinematics is completely distinct from
that of [Fe\,{\sc ii}] and Br$\gamma$, showing radial expansion in
an off-centered $\sim$\,100\,pc radius ring in the galaxy plane.
The expansion velocity decelerates from $\approx$\,200\,km\,s$^{-1}$
in the inner border of the ring to approximately zero at the outer
border where our previous studies found a 10 Myr stellar population.

\end{abstract}


\section{Introduction}

The kinematics of the narrow line region (NLR) of NGC\,1068 has been
the subject of many studies over the years, the most recent and
detailed being those of \citet{cecil02} and \citet{das06}, based on
long-slit optical spectroscopy of the inner few hundred parsecs
obtained with the Space Telescope Imaging Spectrograph aboard the
Hubble Space Telescope ({\it HST-STIS}).
\citet{das06} have shown that the NLR gas is outflowing in a hollow
bi-cone, along which the gas seems to be accelerated up to 140\,pc
from the nucleus and then decelerated.

Integral field optical spectroscopy of the inner 1.5\,kpc was
obtained by \citet{emsellem06} using the instrument SAURON -- at
lower spatial resolution but covering a larger region of the galaxy
than in \citet{das06}.
\citet{emsellem06} observed the outflow in [O\,III]~$\lambda$\,5007
and H$\beta$, finding also streaming motions towards the nucleus and
emission knots of H$\beta$ attributed to regions of star formation
in the galaxy plane.
\citet{gerssen06} presented optical integral field observations of
the inner 400\,pc using the Gemini Multi-Object Spectrograph
Integral Field Unit (GMOS-IFU) at a better spatial resolution
($\approx$\,40\,pc) which revealed, besides high-velocity gas
emission from  outflowing gas, also some emission from the gas in
the galaxy disc. 

In the near-infrared (hereafter near-IR), \citet{muller09} and
\citet{muller11} have mapped the gas kinematics via integral field
observations using the instrument SINFONI at the Very Large
Telescope (VLT).
While \citet{muller09} focus on the H$_2$ kinematics in the inner
few tens of pc which revealed inflows, \citet{muller11} focus on the
Br\,$\gamma$ and coronal lines kinematics, which are dominated by
outflows.

In \citet{riffel13} (hereafter Paper~I) we have used the Gemini
Near-infrared Integral Field Spectrograph (NIFS) on the Gemini North
telescope operating with the adaptive optics module ALTAIR to map
the flux distributions and excitation of the gas within the inner
$300$\,pc, while in \citet{storchi12} we have studied the stellar
population.

In this paper, we present detailed measurements of the gas
kinematics observed in the near-IR emission lines
[Fe\,II]~$\lambda\,1.644\,\mu$m, H$_2\,\lambda\,2.122\,\mu$m and
Br$\gamma$ using channel maps and position velocity diagrams.
We concluded this is the best way to separate the multiple kinematic
components which are present in the emitting gas in this galaxy.
The [Fe\,II] gas kinematics shows a ``broader" outflow than
previously observed in [O\,III] and in the near-IR coronal lines,
being similar to that observed in planetary nebulae and consistent
with an origin in an accretion disc wind \citep{proga04} or a torus
wind \citep{elvis00}.
We test three models with different geometries in order to reproduce
the outflow.

Throughout this paper we adopt a distance of 14.4\,Mpc to NGC\,1068
which correspond to a scale of 72\,pc\,arcsec$^{-1}$ at the galaxy.
This paper is organized as follows.
In Sec. 2 we present the observations, in Sec. 3 we discuss previous
studies relevant to the present work, in Sec. 4 we present the gas
kinematics (channel maps and {\it p-v} diagrams) , in Sec. 5 we give
our interpretation to the kinematics of the [Fe\,{\sc ii}],
Br$\gamma$ and H$_2$ , in Sec. 6 we perform the modeling and
analysis, in Sec. 7 we calculate the mass outflow rate and
corresponding power and in Sec. 8 we present our conclusions.

\section{Observations and Data}

The spectroscopic data was obtained at the Gemini North telescope,
with NIFS (see \citealt{mcgregor03}), operating with the ALTAIR
adaptive optics module.
The data were obtained on the nights of November 26 and December 03
and 09, 2008, and in the present paper we have used spectra obtained
in the H and K spectral bands at a resolving power of 5290.

NIFS has a square field of view of 3\arcsec\ $\times$ 3\arcsec\ and
was set to a position angle of 300 degrees for the observations,
resulting in a spatial sampling of 0.044 arcsec perpendicular to the
jet (vertical direction in our images) and 0.103 arcsec along it
(horizontal direction in our images).
The angular resolution of the data, measured as the FWHM of the
spatial profile of a star is $0\farcs11\,\pm\,0\farcs02$ at the H
and K bands, corresponding to $\cong 8\,$pc at the galaxy.

Dithering was applied in order to cover a field-of-view of $\cong$
5\arcsec\ $\times$ 5\arcsec.
Additional details about the data can be found in Paper~I .

\section{Previous works}

\begin{figure}
  \includegraphics[width=0.5\textwidth]{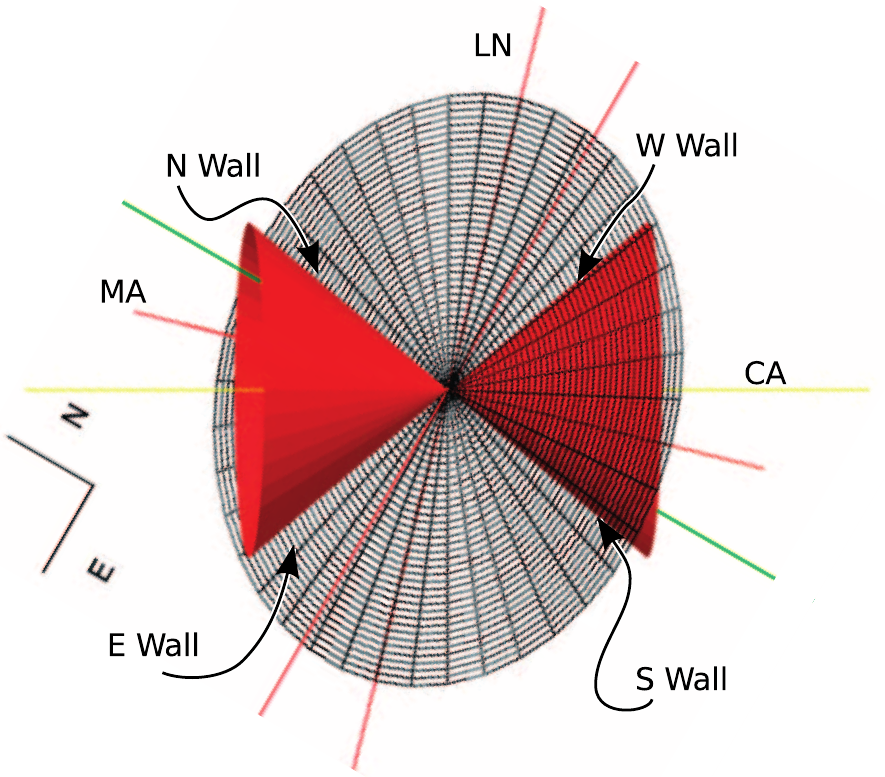}
  \caption{ Geometry of the outflow relative to the galactic disc,
  adapted from \citet{das06}, after a rotation to
  match the orientation of our data.
  The walls of the cone are labeled as N, E, S and W Walls.
  The labels LN, MA and CA correspond to the line of nodes, minor
  axis and cone axis, respectively.}
  \label{fig-das-model}
\end{figure}

On the basis of {\it HST-STIS} spectra, \citet{das06} proposed a
kinematic model in which the NLR has the geometry of a hollow bicone
oriented at PA of $30^\circ\!\pm 2$ as illustrated in Figure
\ref{fig-das-model}.
This figure was adapted from their Figure 10 and rotated to the same
orientation as our data.
We have identified in this figure the line of nodes (LN), minor axis
(MA) and cone axis (CA) as derived by \citet{das06}.
The walls of the cone -- corresponding to the areas intersecting
perpendicularly the plane of the sky -- were also labeled in this
figure as N, E, S and W Walls.
The galactic plane is inclined by $40^\circ$ such that the far side
of the galaxy is the NE \citep{dev91}.
The NE side of the cone is in front of the galactic disc, while  the
SW side is behind and apparently obscured by the galactic disc.
The bicone axis is tilted by $5^\circ$ relative to the plane of the
sky such that the NE side is oriented (slightly) toward us, while
the SW side is oriented away from us.
The cone is hollow but has ``thick" walls, corresponding to inner
and outer half-opening angles of $20^\circ$ and $40^\circ$.
The NLR gas flows away from the nucleus along the cone, reaching a
maximum velocity of 2000~km~s$^{-1}$ at 140\,pc from the nucleus.

\begin{figure}
  \includegraphics[width=\linewidth]{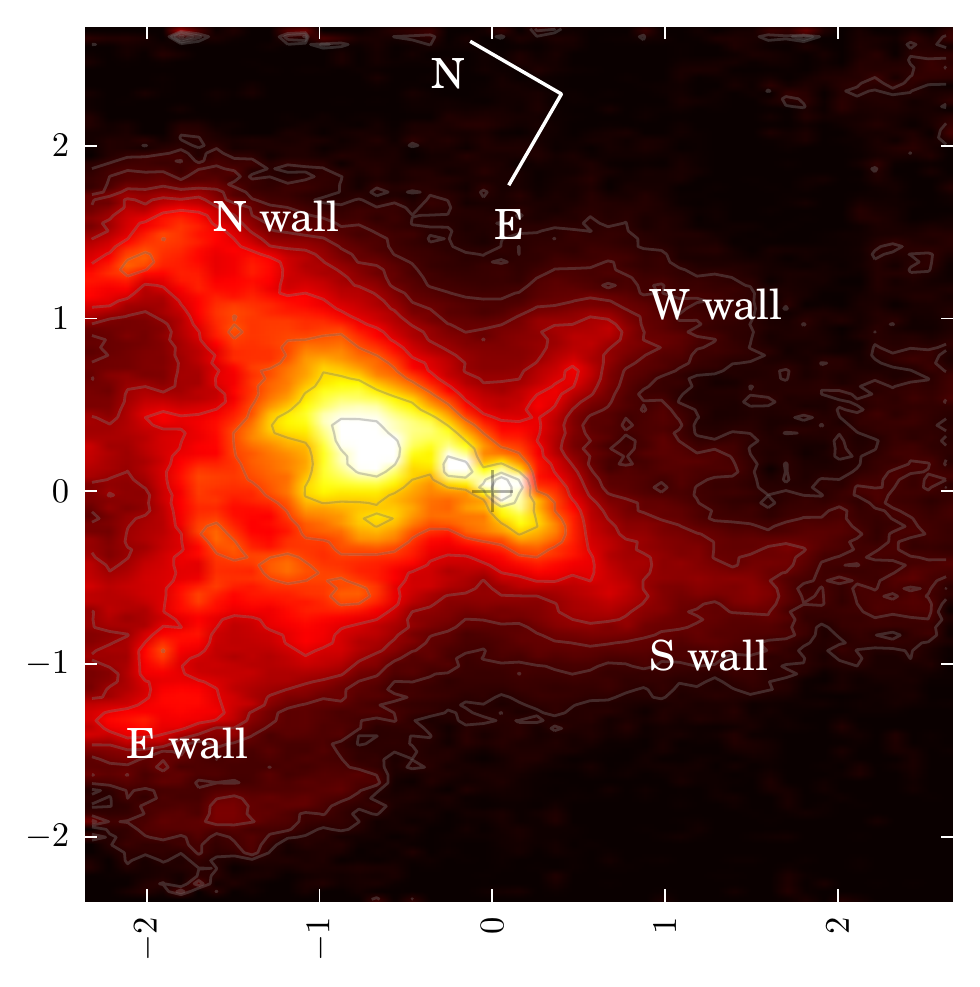}
  \caption{Integrated flux map in the [Fe\,{\sc
  ii}]\,$\lambda\,1.644\,\mu$m emission line.
  Note the hourglass (rather than a conical) shape of this emission.
  The north, south, east and west walls are indicated by the labels.
  Spatial units are in arcsec and the position of the nucleus
  $(0,0)$ is indicated by a plus sign.
  The orientation is indicated by the arrows.
  Contours are used to best delineate the structures.
  }
  \label{fig-flux-map-fe}
\end{figure}

\begin{figure}
  \includegraphics[width=\linewidth]{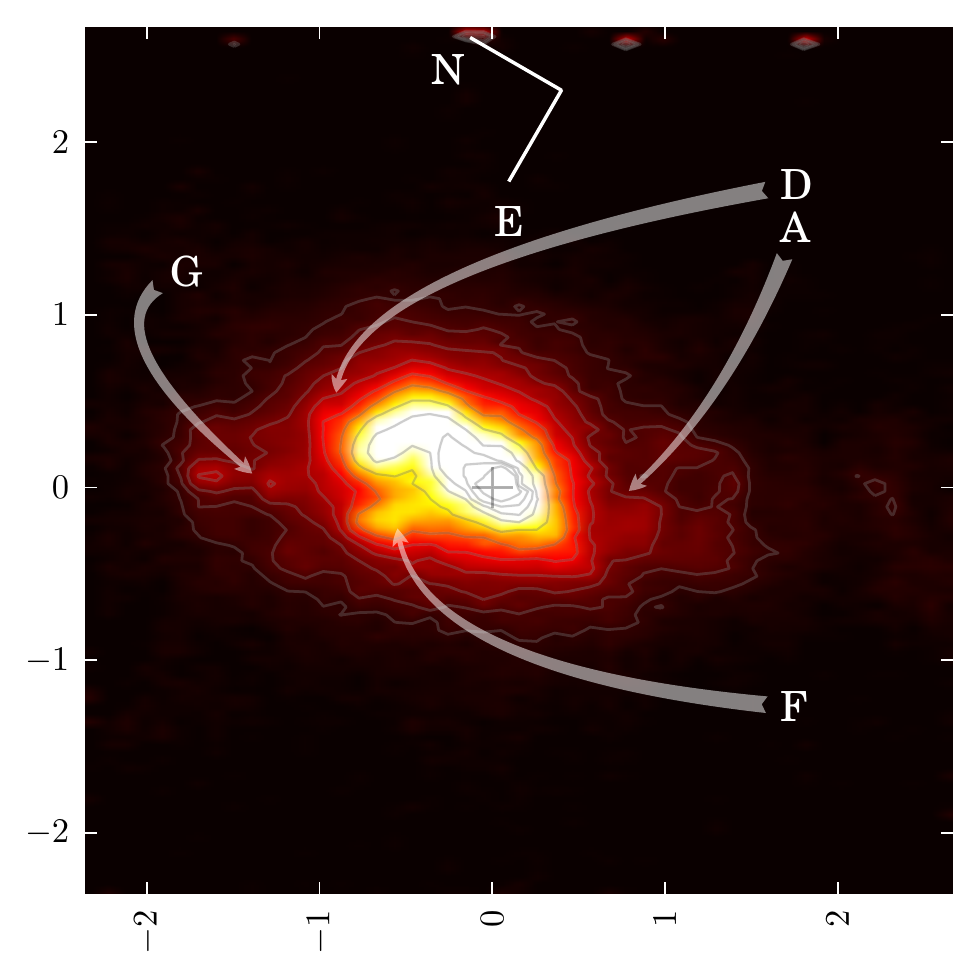}
  \caption{
  Integrated flux map in the Br$\gamma$ line.
  Details are the same as in Fig. \ref{fig-flux-map-fe}, otherwise
  indicated bellow.
  The letters A, D, F and G label structures corresponding to those
  previously identified in the [O\,{\sc iii}] flux map from
  \citet{evans1991}.
  }
  \label{fig-flux-map-bra}
\end{figure}

\begin{figure}
  \includegraphics[width=\linewidth]{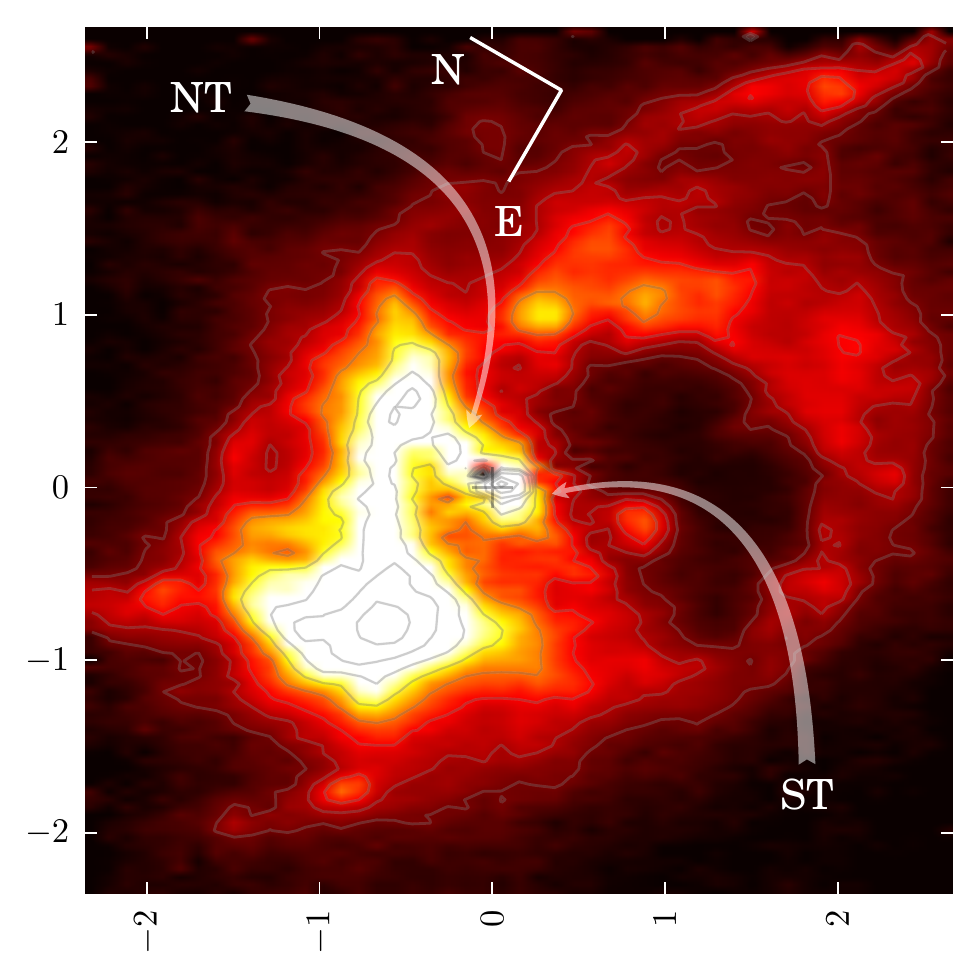}
  \caption{
  Integrated flux map in the H$_2$~$2.12\,\mu$m line.
  Details are the same as in Fig. \ref{fig-flux-map-fe}, otherwise
  indicated bellow.
  NT and ST identify the northern and southern ``tongues" from
  \citet{muller09}.
  }
  \label{fig-flux-map-h2}
\end{figure}

We present the integrated flux maps from Paper~I in the lines
[Fe\,{\sc ii}]\,$\lambda\,1.644\,\mu$m, Br$\gamma$ and
H$_2\,\lambda\,2.122\,\mu$m, in Figures \ref{fig-flux-map-fe},
\ref{fig-flux-map-bra} and \ref{fig-flux-map-h2}, respectively.
In Fig. \ref{fig-flux-map-fe} we identify the N, E, S and W walls of
the bicone.
In Fig. \ref{fig-flux-map-bra} the features indicated by arrows
correspond to structures A, D, F and G previously identified in the
[O\ {\sc iii}] map by \citet{evans1991}.
In Fig. \ref{fig-flux-map-h2} the structures identified by
\citet{muller09} as {\it northern tongue} (NT) and {\it southern
tongue} (ST) are also indicated by arrows.
The main feature seen in the H$_2$ flux map is a ring which has its
centre offset from the position of the nucleus by $0\farcs6$ to the
SW \citep{muller09}.
We adopt as the position of the nucleus the location of the maximum
of the K-band continuum emission.

\section{Gas kinematics}

In a previous study, \citet{mazzalay13} has used the same NIFS data
we used in the present paper to map the gas kinematics of the
coronal line region, which is consistent with an origin  of the
emission in the inner part of the NLR.
Here we present and discuss the kinematics of the low ionization
lines of [Fe\,II]~$\lambda\,1.644\,\mu$m,
H$_2\,\lambda\,2.122\,\mu$m and Br$\gamma$, using velocity channel
maps and position-velocity (hereafter {\it p-v}) diagrams.
These two types of maps are two-dimensional cuts through the 3D
integral field data: the channel maps are cuts across the wavelength
(velocity) axis and the {\it p-v} diagrams are cuts across one of
the position axes.

The channel maps are presented in Figures \ref{fig-channel-map-fe}
to \ref{fig-channel-map-h2} and show emission line flux
distributions in velocity channels along the line profile from
blueshifts to redshifts after the subtraction of the systemic
velocity of the galaxy, adopted as $1137$~km~s$^{-1}$
\citep{huchra99}.
Each velocity channel integrates 3 pixels in the velocity axis
corresponding to 87.6~km~s$^{-1}$ for [Fe\,{\sc ii}] and
88.7~km~s$^{-1}$ for Br$\gamma$ and 2 pixels in H$_2$ corresponding
to 60.3~km~s$^{-1}$.
The central velocity $v$ is written at the bottom of each panel.
The central velocities were chosen so that the velocity range covers
all the line profile from blueshifts to redshifts.
In all panels we draw an auxiliary dashed gray line indicating the
orientation of the galactic line of nodes adopted as 80\degr, as
obtained from the observation of the stellar kinematics of the inner
regions by \citet{storchi12}.
The N and E arrows show the orientation of our images in the sky and
the label ``far side'' indicates the far side of the galactic disc.

The {\it p-v} diagrams are presented in Figures \ref{fig-pv-fe} and
\ref{fig-pv-h2} for [Fe\,{\sc ii}] and H$_2$, respectively.
The bottom right panel depicts the integrated flux map of the line
showing the slits of the {\it p-v} diagrams running horizontally in
the figure.
In the {\it p-v} panels the velocity runs along the horizontal axis
(PA~$= 30^\circ$) and the position along the vertical axis with NE
at the bottom (negative positions) and SW at the top (positive
positions).
In the top left of each {\it p-v} panel the central position of the
slits is written in arcsecs.

In the next subsections we describe the channel maps and {\it p-v}
diagrams for the ionized and molecular gas as well as our
interpretation.

\subsection{[Fe\,{\sevensize \it II}]}

\subsubsection{Channel Maps}

\begin{figure*}
  \includegraphics[width=\textwidth]{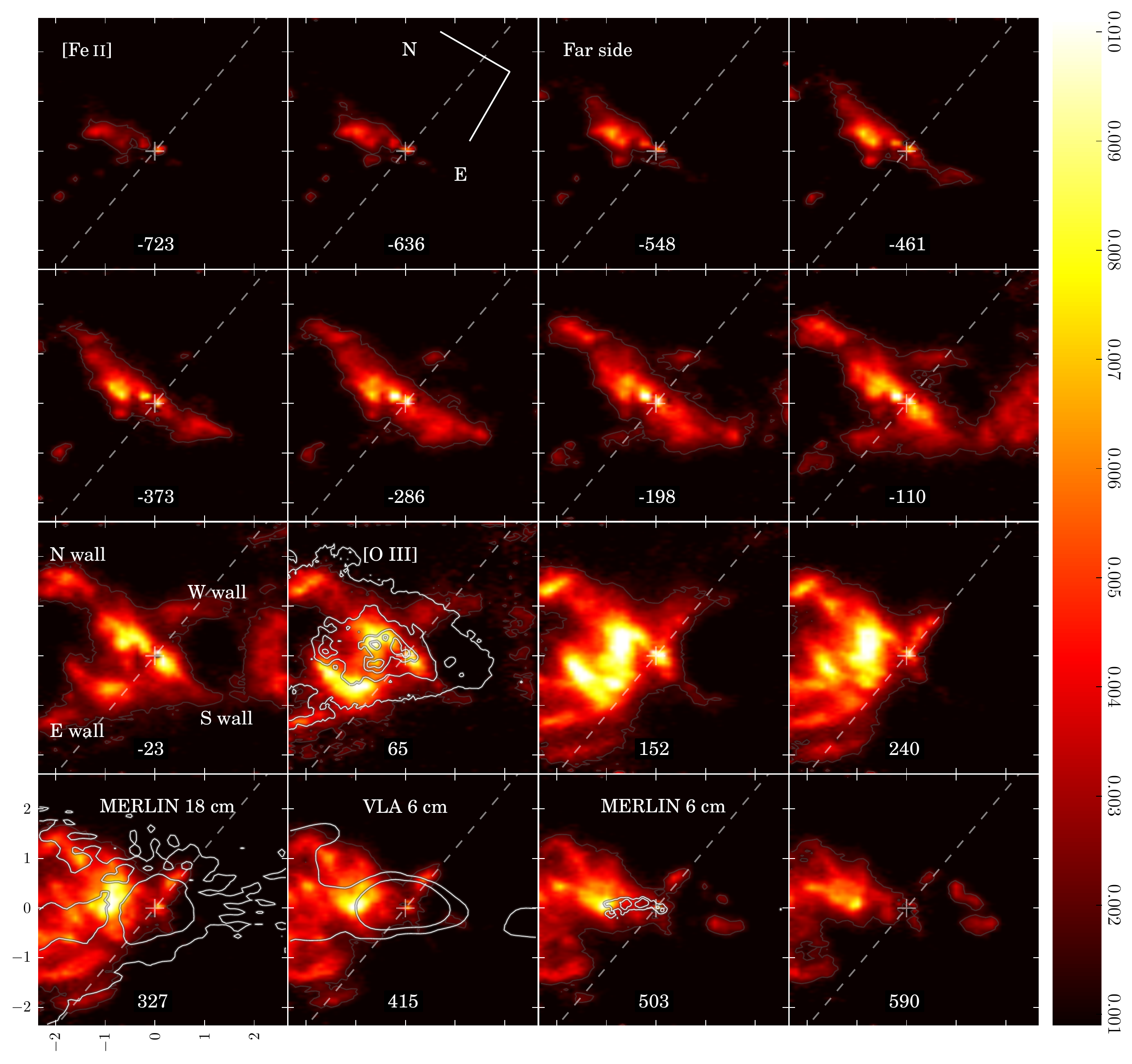}
  \caption{
  Channel maps of the [Fe\,{\sc ii}] line.
  The channel velocities are written in the central bottom side of
  every panel and are given in km~s$^{-1}$.
  The line of nodes is indicated by the dashed gray line and the
  assumed far side of the galaxy is indicated in the panel with
  velocity $-548$~km~s$^{-1}$.
  The orientation is indicated by the arrows in the panel with
  velocity $-636$~km~s$^{-1}$.
  The N, E, S and W walls are indicated in panel with velocity
  $-23$~km~s$^{-1}$
  The radio contours of MERLIN 18~cm, VLA 6~cm and MERLIN 6~cm are
  overlaid, respectively, in the panels with velocities $327$, $415$
  and $503$~km~s$^{-1}$ and [O\,{\sc iii}] contours are shown in the
  panel with velocity $65$~km~s$^{-1}$.
  A color bar on the right gives the correspondence between color
  tones and flux in units of
  10$^{-15}$~erg~cm$^{-2}$~s$^{-1}$~\AA$^{-1}$.
  }
  \label{fig-channel-map-fe}
\end{figure*}

The [Fe\,{\sc ii}] channel maps are shown in Fig.
\ref{fig-channel-map-fe}, where we see emission in velocity channels
from $-723$~km~s$^{-1}$ to $590$~km~s$^{-1}$.
In fact we found [Fe\,{\sc ii}] emission in velocity channels
ranging from about $-1500$~km~s$^{-1}$ to $770$~km~s$^{-1}$.
The channels not shown in the figure were omitted because they
present a structure that does not change much from what is seen in
the first and last channels.
The emission seems to extend beyond the instrument field of view in
the majority of the panels.
The central and redshifted panels delineate a ``bowl" shape which
can also be described as similar to an hourglass extending along the
horizontal direction in our images, from NE to SW with axis along
PA~$\sim30\degr$.
This hourglass seems to correspond to the walls of the bipolar
outflow modelled by \citep{das06} with a hollow cone shape.
The walls of this structure are oriented approximately along N and E
to one side of the nucleus and along S and W to the other side.

The highest blueshift structures detached from the nucleus are
observed in the NE outflow up to a velocity channel centered at 
$-1249$~km~s$^{-1}$ (not presented in the figure).

The N wall of the hourglass is observed covering a range of
blueshifts and redshifts, while the E wall is observed mostly in
redshift.
The N and E walls of the hourglass do not keep the same PA and
structure over the velocity channels due to the patchy nature of the
emitting gas and also because of the hourglass geometry and its
relative orientation to the line of sight.
The opening angle of the walls is widest ($\sim$\,$100^\circ$) at
the low redshift channels.

The S wall is observed mostly in blueshift, while the W wall is the
faintest, observed in the lowest velocity channels.

The most luminous part of the outflow is the ``fan-like" structure 
observed to the NE in the redshift channels, corresponding to the back 
wall of the NE outflow.

Radio MERLIN 18~cm, VLA 6~cm and MERLIN 6~cm contours are
superimposed, respectively, to the panels with velocities
$327$~km~s$^{-1}$, $415$~km~s$^{-1}$ and $503$~km~s$^{-1}$.
The MERLIN 18~cm emission fills the region between the N and E walls
of the hourglass and also extends towards SW partially filling that
side of the hourglass.
The MERLIN 6~cm is very compact and extends towards NE at
PA~$\sim30\degr$ along the axis of the hourglass.
The VLA 6~cm radio image is more extended than the MERLIN 6~cm.
Its emission partially fills the NE side of the hourglass and has a
detached patch of emission to the SW.

In the velocity channel map at $65$~km~s$^{-1}$, the contours of the
{\it HST} [O\,{\sc iii}] integrated flux image are shown.
The emission covers most of the field and extends toward NE along
PA~$\sim30\degr$.
There is good correspondence between [O\,{\sc iii}] and [Fe\,{\sc
ii}] flux distributions within 1 arcsec from the nucleus between N
and NE at the highest flux levels.
At lower flux levels the correspondence is poorer and in the
[O\,{\sc iii}] contours we do not see the ``bowl''shape seen in the
[Fe\,{\sc ii}] flux distributions of the redshift channels.

\subsubsection{Position Velocity Diagrams}

\begin{figure*}
  \includegraphics[width=\textwidth]{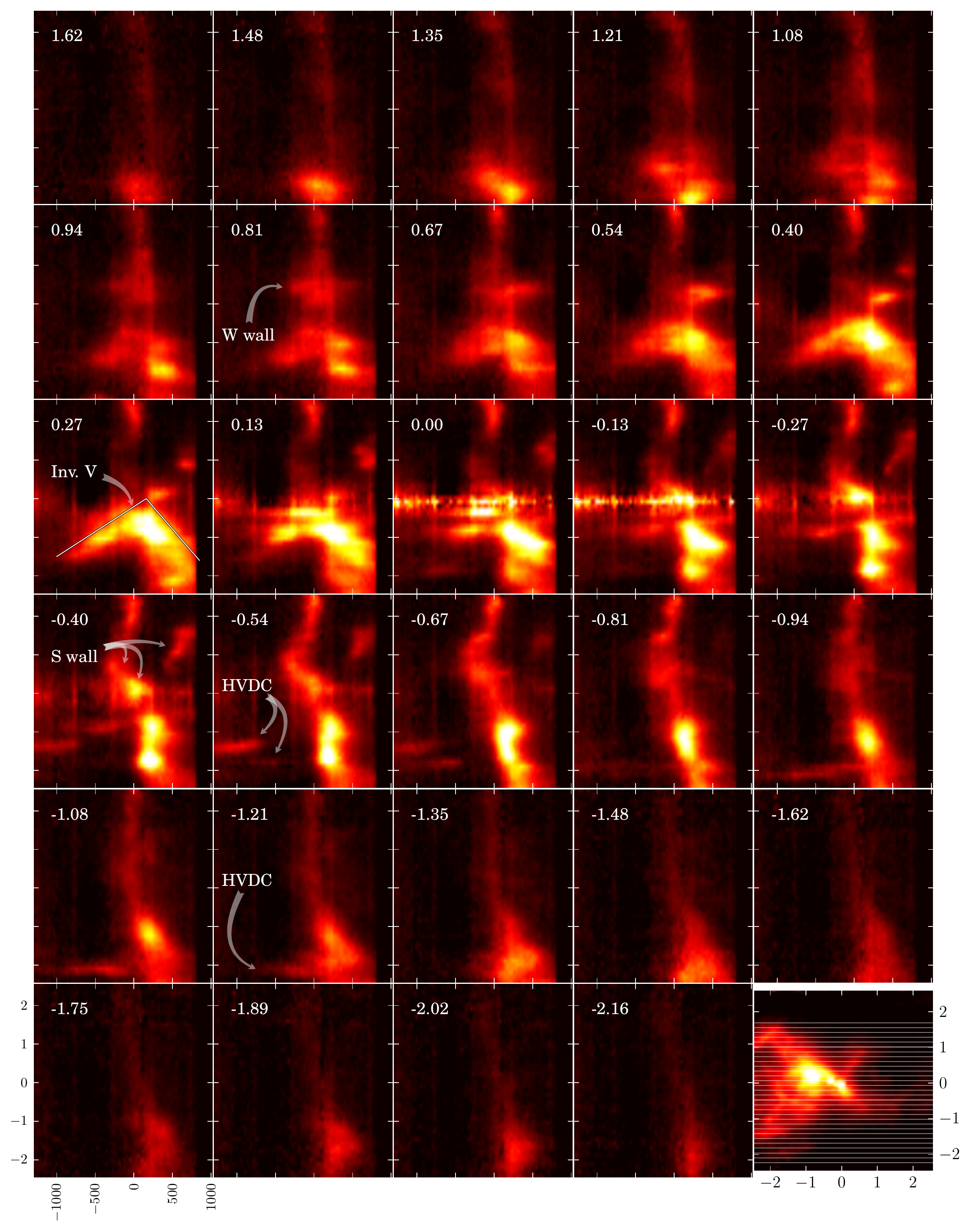}
  \caption{
  {\it p-v} diagram of the [Fe\,{\sc ii}] line.
  The numbers in the panels correspond to the slit centre position
  relative to the nucleus along the `y'~axis of our data.
  The slits are indicated by the horizontal solid lines overlaid in
  the [Fe\,{\sc ii}] integrated flux map shown in the lower right
  panel.
  Vertical stripes seen at velocities $-750$~km~s$^{-1}$,
  $-250$~km~s$^{-1}$, $250$~km~s$^{-1}$ and $750$~km~s$^{-1}$ are
  artifacts.
  }
  \label{fig-pv-fe}
\end{figure*}

Fig. \ref{fig-pv-fe} presents the [Fe\,{\sc ii}] {\it p-v} diagrams
in panels corresponding to adjacent slits 0\farcs 13 wide.
The lower right panel of the figure presents the integrated
[Fe\,{\sc ii}] emission line flux map.
Despite our efforts to clean all artifacts from the data, regularly
spaced vertical lines resulting from the continuum subtraction
appear in velocities $-750$~km~s$^{-1}$, $-250$~km~s$^{-1}$,
$250$~km~s$^{-1}$ and $750$~km~s$^{-1}$ mainly in the central panels
and should not be confused with real structures by the reader.

The {\it p-v} flux distribution is dominated by emission from the NE
side of the outflow in most panels (lower half of the panels)
although weak emission from the galactic disc is also seen running
vertically at $v = 0$~km~s$^{-1}$.
The most prominent feature is an inverted ``v''-shaped structure
which appears in slits ranging from 1\farcs08 to 0\farcs00 therefore
coming mainly from the N side of the hourglass structure.
This structure is indicated in the panel corresponding to the slit
$0\farcs27$ by solid straight lines and the label ``Inv. V''.
We observe that this feature is asymmetric with the blueshift side
reaching almost twice as much velocity as the redshift side for
equivalent distance from the nucleus.

We have also identified in Fig. \ref{fig-pv-fe} the emission
coming from the W wall -- observed in the slits 0\farcs 94 to
0\farcs 13 (label ``W wall''), and that coming
from the S wall -- observed in the slit positions 0\farcs00 to
-0\farcs81 (label ``S wall'').

In the slits corresponding to negative positions up to $0\farcs 00$
we see the kinematic signatures of many individual high velocity
dispersion clouds (labeled HVDC) identifiable as horizontally
elongated emitting areas.
The velocity dispersion of these structures range from $\sim
300$~km~s$^{-1}$ to $\sim 900$~km~s$^{-1}$.
Each cloud appears in two or three adjacent panels indicating they
have angular sizes ranging from 0\farcs 2 to 0\farcs 4.
These clouds originate in the front wall of the NE side of the
hourglass.

\subsection{Br$\gamma$: Channel Maps}

\begin{figure*}
  \includegraphics[width=\textwidth]{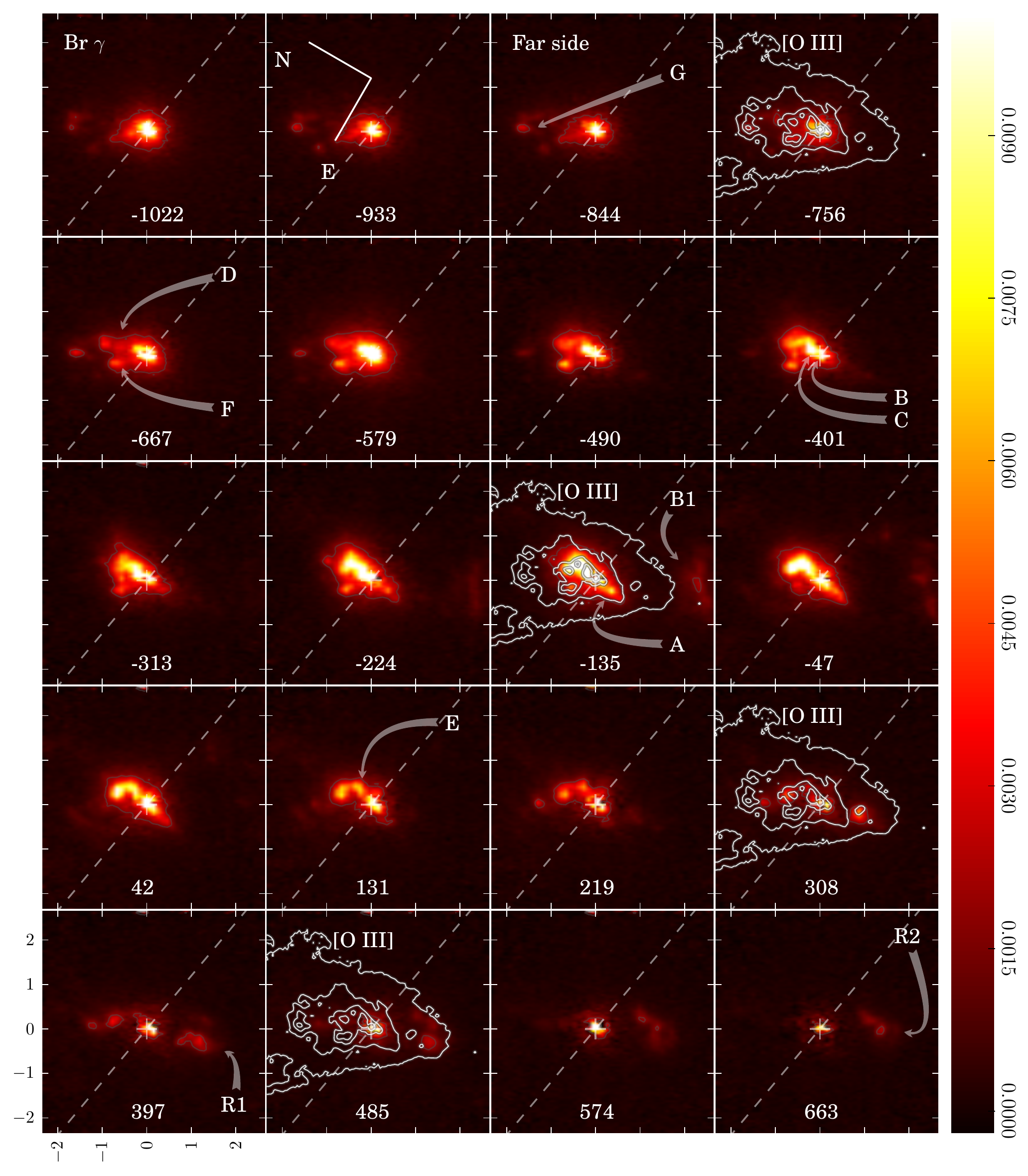}
  \caption{
  Channel maps of the Br$\gamma$ line.
  The velocities are written in the central bottom part of every
  panel and are given in km~s$^{-1}$.
  The major axis is indicated by the dashed gray line and the
  assumed far side of the galaxy is indicated in the panel with
  velocity $-844$~km~s$^{-1}$.
  The orientation is indicated by the arrows in the panel with
  velocity $-933$~km~s$^{-1}$.
  The [O\,{\sc iii}] contours are overlaid in the panels with
  velocities $-756$, $-135$, $308$ and $485$~km~s$^{-1}$.
  A color bar on the right gives the correspondence between color
  tones and flux in units of
  10$^{-15}$~erg~cm$^{-2}$~s$^{-1}$~\AA$^{-1}$.
  }
  \label{fig-channel-map-bra}
\end{figure*}

The Br$\gamma$ channel maps are shown in Fig.
\ref{fig-channel-map-bra}, where we see emission in velocity
channels ranging from $-1022$~km~s$^{-1}$ to $663$~km~s$^{-1}$.
We, however, found Br$\gamma$ emission in velocity channels ranging
from about $-1110$~km~s$^{-1}$ to redshifts beyond
$1000$~km~s$^{-1}$.
The omitted channels beyond redshift $663$~km~s$^{-1}$ show only the
central component getting fainter with redshift and the channel
$-1110$~km~s$^{-1}$ shows a faint similar structure to that seen in
the channel $-1022$~km~s$^{-1}$.

The emission structure is more compact than that seen in the
[Fe\,{\sc ii}] channels and is composed mainly by emission to the
N-NE seen in both blueshifted and redshifted channels and to the SW
in channels from $-331$~km~s$^{-1}$ to $663$~km~s$^{-1}$.

The contours of the {\it HST} [O\,{\sc iii}] line emission are
overplotted on four channels corresponding to the velocities
$-756$~km~s$^{-1}$, $-135$~km~s$^{-1}$, $308$~km~s$^{-1}$ and
$485$~km~s$^{-1}$.
These are the same contours as those overplotted in the [Fe\,{\sc
ii}] channel maps.
The correlation between the [O\,{\sc iii}] contours and Br$\gamma$
flux distribution is better than that between [O\,{\sc iii}] and
[Fe\,{\sc ii}].
Almost every Br$\gamma$ feature (covering the whole velocity range)
corresponds to an [O\,{\sc iii}] feature.
In particular we identify \citep[from][]{evans1991} in Fig.
\ref{fig-channel-map-bra} the components A, C, D, F and G which are
observed in blueshift, component E observed in blueshift and
redshift and component B, the nuclear component, observed in all
velocity channels.

We have also identified some new structures in the SW side of the
bipolar outflow which we call B1, R1 and R2.
B1 is observed in the velocity channels from $-224$~km~s$^{-1}$ to
$-47$~km~s$^{-1}$ as an area of faint Br$\gamma$ emission indicated
in panel of $v = -135$~km~s$^{-1}$. 
This area has a SE-NW extension of about $2\arcsec$ and is observed
at $\sim 2$~arcsec SW of the nucleus, although it seems to extend to
SW beyond the field of view.
We attribute this emission to the front wall of the SW outflow.
R1 and R2 are two redshift components (shown in panels of $v =
397$~km~s$^{-1}$ and $663$~km~s$^{-1}$ respectively).
R1 is observed in the velocity range $131$~km~s$^{-1}$ to
$574$~km~s$^{-1}$ at $\sim 1\farcs2$ SW of the nucleus at
PA~$=193^\circ$ and R2 in the velocity range $397$~km~s$^{-1}$ to
$751$~km~s$^{-1}$ distant from the nucleus by $\sim 1\farcs2$ at
PA~$\sim210^\circ$.
We atribute the origin of R1 and R2 to the back wall of the SW
outflow.
Direct comparison of the [O\,{\sc iii}] contours to the Br$\gamma$
flux map at the positions of R1 and R2 reveals a correspondence of
[O\,{\sc iii}] emission to the Br$\gamma$ components.
These redshift components are also observed in [Fe\,{\sc ii}] in
channels from $327$~km~s$^{-1}$ to $678$~km~s$^{-1}$ for R1 and in
channels from $503$~km~s$^{-1}$ to $678$~km~s$^{-1}$ for R2.
R1 and R2 are, respectively, in vertical coordinates $-0.3$ and
$0.1$ arcsec approximately and, therefore, are separated by $\simeq
0.4$ arcsec.

\subsection{Interpretation: Ionized gas kinematics}

\label{subsec-ionized-gas}

The ionized gas kinematics has been previously modeled by
\citet{das06} as a bi-conical outflow with axis at PA~$= 30^\circ$
in which the front walls produce the blueshifted emission and back
walls, the redshifted emission.
Their model was fitted to [O\,{\sc iii}] kinematic data measured
using HST STIS long-slit spectroscopy.
Fig. \ref{fig-channel-map-bra} shows that the flux distribution in
[O\,{\sc iii}] (contours) is very similar to that of Br$\gamma$ and
so we assume that their emission originate in the same gas and
consequently the kinematics of the [O\,{\sc iii}] and Br$\gamma$
should also be similar.
The blueshift channels correspond to the front wall of the outflow
and the redshift channels correspond to the back wall.
From this figure it is not clear that the geometry of the outflow is
conical as derived by \citet{das06}.
In our data the brightest emission comes, from a narrow region
extending from the nucleus to the N-NE, being stronger in the
blueshift channels.
This emission comes from the front wall of the outflow which is at
higher galactic latitude than the emission from the back wall, that
is observed in redshift.
Additionally, the emission coming from the SW side of the outflow is
observed in redshift corresponding to emission from the back wall of
the outflow which is at higher galactic latitude than that of the
front wall.

\citet{das06} derived the conical geometry for the outflow from the
kinematics that they could obtain from deep STIS long-slit
spectroscopy of the whole region, not from the flux maps.

The approximately conical geometry is better seen in the [Fe\,{\sc
ii}] channel maps (Fig. \ref{fig-channel-map-fe}), and in
particular, in the redshift channels.
While the Br$\gamma$ emission distribution is concentrated and
mostly directed towards north, the [Fe\,{\sc ii}] emission
distribution encompasses it and extends further spanning a wide
hourglass shaped area more clearly seen to NE of the nucleus between
PA~$=0^\circ$ and $90^\circ$ in the redshift channels.

The difference between [Fe\,{\sc ii}] and Br$\gamma$ flux
distribution and kinematics can be understood as follows.
As discussed by \citet{mouri00}, most of the [Fe II] emission comes
from a region where H is only partially ionized. 
In the completely ionized region, as the ionization potential of Fe+
is just 16.2\,eV, Fe+ is further ionized and there is not much
[Fe\,{\sc ii}] emission. 
Extensive partially ionized zones around the NLR of Seyfert galaxies
are created by X-rays that have small cross sections and penetrate
deeper in the gas than less energetic photons, heating the gas.
The [Fe\,{\sc ii}]\,$\lambda\,1.644\,\mu$m emission is excited in
these partially ionized zones by collisions with electrons, as well
as by shocks that may be present due to the interaction of AGN jets
and winds with the ambient gas.
The extent of the partially ionized region determines where we may
find [Fe\,{\sc ii}] emission.
The larger extent of the [Fe\,{\sc ii}] emitting region compared to
that of the Br$\gamma$ can thus be understood as caused by the
existence of a partially ionized zone extending beyond the fully
ionized hydrogen zone where Br$\gamma$ is produced.

The enhancement of the [Fe\,{\sc ii}] emission to the NE in the
redshift channels when compared with the Br$\gamma$ emission can
also be attributed to the interaction of the back wall of the
outflow with the far side of the galaxy disc.
The shocks produced in this interaction could provide the energy for
the excitation of [Fe\,{\sc ii}], besides some contribution from
collisions with electrons in the gas heated by X-rays from the AGN.
Support for the presence of shocks is provided by line ratio maps
presented in Paper I.
Towards the borders of the line flux distributions, in particular to
the NE border, there is an increase of the [Fe\,{\sc ii}]/Pa$\beta$
line ratio to values up to $\sim$\,4, consistent with shock
excitation.

We conclude that the [Fe\,{\sc ii}] emission can be observed beyond
the region where  [O\,{\sc iii}]  and H recombination lines can be
observed, providing a better coverage of the outflow.

\subsection{H$_2$}

\subsubsection{Channel Maps}

\begin{figure*}
  \includegraphics[width=\textwidth]{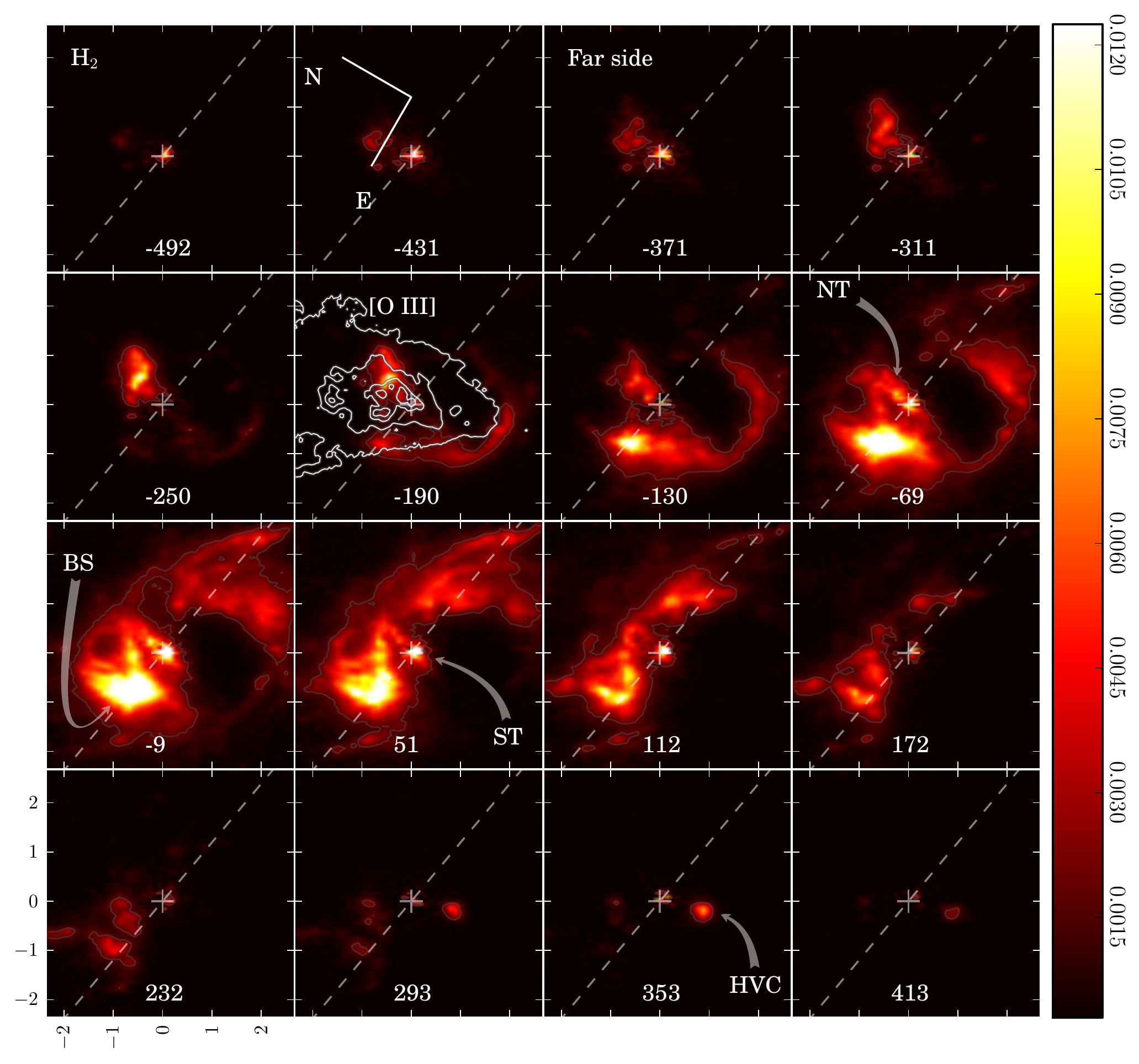}
  \caption{
  Channel maps of the H$_2$ line.
  The velocities are written in the central bottom part of every
  panel and are given in km~s$^{-1}$.
  The major axis is indicated by the dashed gray line and the
  assumed far side of the galaxy is indicated in the panel with
  velocity $-371$~km~s$^{-1}$.
  The orientation is indicated by the arrows in the panel with
  velocity $-431$~km~s$^{-1}$.
  The [O\,{\sc iii}] contours are overlaid in the panels with
  velocity $-190$~km~s$^{-1}$.
  A color bar on the right gives the correspondence between color 
  tones and flux in units of
  10$^{-15}$~erg~cm$^{-2}$~s$^{-1}$~\AA$^{-1}$.
  }
  \label{fig-channel-map-h2}
\end{figure*}

The H$_2$ emission is detectable mainly in the velocity range $-492$
to $413$~km~s$^{-1}$ as can be seen in Fig.
\ref{fig-channel-map-h2}.
The distribution of H$_2$ flux emission along the velocity channels
is very different from that of the [Fe\,{\sc ii}] and Br$\gamma$,
indicating that they are tracing distinct gas components and
processes.
The strongest emission is detected in the central channels (around
zero velocity) and exhibits a ring like morphology (see Paper~I and
\citealt{muller09}).

The emission of the ring in redshift appears in the velocity
channels from $51$~km~s$^{-1}$ to $232$~km~s$^{-1}$ and it comes
from the N half of the ring while in the blueshift channels from
$-250$~km~s$^{-1}$ to $-69$~km~s$^{-1}$ the emission comes from the
S half of the ring.

In the highest blueshift velocity channel we observe nuclear
emission as well as a clump of faint emission distant from the
nucleus by $\sim 1\arcsec$ at PA~$= 10^\circ$.
The clump is the tip of an elongated emission structure seen in the
channels with velocities $-431$~km~s$^{-1}$ to $51$~km~s$^{-1}$,
which include the {\it northern tongue} (NT as in
\citealt{muller09}) most clearly seen from channel $v =
-69$~km~s$^{-1}$ to channel $v = 51$~km~s$^{-1}$.

The {\it southern tongue} (ST as in \citealt{muller09}) is
identified in channel $v = 51$~km~s$^{-1}$ and observable in
channels from $v = -69$~km~s$^{-1}$ to $51$~km~s$^{-1}$.

Contours of the {\it HST} [O\,{\sc iii}] line emission are
overplotted on the velocity channel $-190$~km~s$^{-1}$.
A comparison of the [O\,{\sc iii}] contours to the H$_2$ maps shows
that the H$_2$ emission knots extending to the N of the nucleus in
the blueshift channels also correspond to emission knots in [O\,{\sc
iii}] although the most extended emission is not correlated.

From channel $-190$~km~s$^{-1}$ to $293$~km~s$^{-1}$ we observe the
brightest structure, indicated in the channel $-9$~km~s$^{-1}$ as BS
(for bright spot).
This structure appears in our data blended to the ring at NE.

In the highest redshift velocity channel we find faint nuclear
emission as well as a clump of emission distant from the nucleus by
$\sim 0\farcs9$ at PA~$= 190^\circ$.
This redshift clump is identified as HVC (for high velocity cloud)
in the velocity channel $353$~km~s$^{-1}$ and is observed in
channels from $v = 293$~km~s$^{-1}$ to $413$~km~s$^{-1}$.
This clump is almost symmetrical to the clump seen in the highest
blueshift channels suggesting they could be redshift and blueshift
counterparts of the same gas ejection event.

\subsubsection{Position Velocity Diagrams}

\label{pv_H2}

\begin{figure*}
  \includegraphics[width=\textwidth]{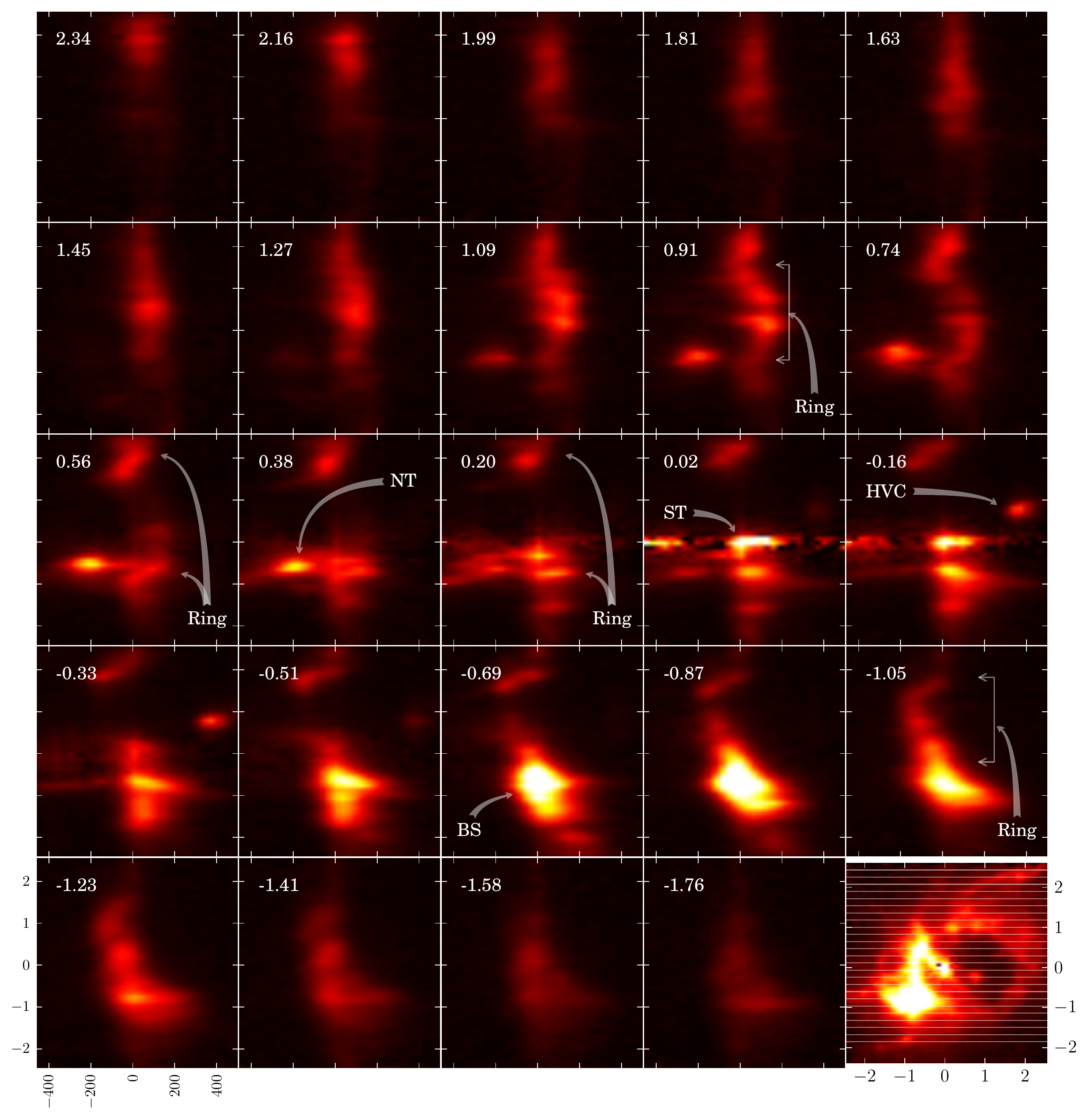}
  \caption{
  {\it p-v} diagram of the H$_2$ line.
  The numbers in the panels correspond to the slit position relative
  to the nucleus along the `y'~axis of our data.
  The slits are indicated by the horizontal solid lines overlaid in
  the H$_2$ integrated flux map shown in the lower right panel.
  }
  \label{fig-pv-h2}
\end{figure*}

The H$_2$ {\it p-v} diagrams are shown in Fig. \ref{fig-pv-h2}.
The integrated H$_2$ emission line flux map is shown in the lower
right panel.

The most conspicuous structures in the integrated flux map are the
nucleus, the ring -- the NE side being its brightest area, the
northern and southern tongues (NT and ST), a bright area to the E of
the ring which we call Bright Spot (BS)  and an isolated cloud
inside the ring, which we call High Velocity Cloud (HVC).
It is interesting to note that neither the nucleus nor the HCV are
at the centre of the ring.

We describe bellow the structures seen in each {\it p-v} panel, from
the bottom slits (SE) to the top slits (NW).  
In the panel $-1\farcs76$, there is a high velocity dispersion
(hereafter $\sigma$) knot at position $-1\arcsec$ with velocities
ranging from $\sim -100$~km~s$^{-1}$ to $\sim 300$~km~s$^{-1}$.
The remaining emission in this slit comes from the galactic plane
and show velocities around zero, with $\sigma \sim
100--200$~km~s$^{-1}$.
The subsequent slits $-1\farcs58$ and $-1\farcs41$ show similar
structures.

From $-1\farcs23$ to $-0\farcs51$ the bright knot of emission at
$\approx\,-0\farcs8$ corresponds to the BS, showing velocities from
zero to $300$~km~s$^{-1}$.

At positive positions, the slits $-1\farcs23$ and $-1\farcs05$ are
tangent to the SE side of the ring and its emission has negative
velocities reaching up to $\sim -200$~km~s$^{-1}$.
The emission in this area has a bow figure: around position
$0\farcs8$, the velocities are more negative than around position
$1\farcs5$ and $0\farcs 0$ meaning that as one moves from the inner
to the outer border of the ring the velocities change from
blueshifts to zero.
In the subsequent slits from $-0\farcs87$ to $0\farcs56$,
the SW side of the ring shows velocities ranging from $\sim
-200$~km~s$^{-1}$ at the inner border to zero at the outer border. 

In the slits $-0\farcs33$ and $-0\farcs16$ we observe the HVC with a
velocity of $\sim 400$~km~s$^{-1}$ and
$\sigma\approx100$~km~s$^{-1}$. 

In the slits from $-0\farcs16$ to $0\farcs74$ the NE side of the
ring shows a similar effect to that described above: a velocity
gradient across the ring, with the inner part of the ring at
position $\sim -0\farcs5$ showing redshifts of up to $\sim
200$~km~s$^{-1}$ and the outer part of the ring showing zero
velocities.

In the slits $0\farcs02$ to $1\farcs38$ the NT appears at position
$\sim -0\farcs6$ with blueshifts of up to $\sim -300$~km~s$^{-1}$,
and seems to be a structure detached from the ring and from the
nucleus.

Finally, the slits $0\farcs74$ to $1\farcs27$ intercept tangentially
the NW side of the ring, showing redshifts of $150$~km~s$^{-1}$ in
the inner border of the ring and velocities close to zero in the
outer border of the ring.

In summary, a general trend is that the NE side of the ring shows
mostly redshifts, while the SW side (which is less luminous) shows
blueshifts indicating expansion of the ring.
But we, in addition, observe a velocity gradient from velocities in
the range $100$ -- $200$~km~s$^{-1}$ observed in the inner border of
the ring to approximately zero velocities in the outer border of the
ring.

\subsection{Interpretation: Molecular gas kinematics}

The H$_2$ kinematics is completely distinct from that of the ionized
gas and implies an origin for the H$_2$ in an expanding ring in the
galaxy plane while the ionized gas is observed in outflow and
extends to high latitudes.
This decoupling between molecular and ionized gas is a result
commonly found in previous studies of the NLR of active galaxies by
our group AGNIFS \citep{barbosa09,sb09,sb10,mrk1066,mrk1157,mrk79}.

Although the morphology of the H$_2$ ring suggests it is a circular
ring in the plane of the galaxy (Paper~I), its kinematics does not
suggest rotation.
In the redshift channels from $51$~km~s$^{-1}$ to $232$~km~s$^{-1}$
the emission comes mostly from the N half of the ring which is in
the far side of the galaxy while in the blueshifted channels from
$-250$~km~s$^{-1}$ to $-69$~km~s$^{-1}$ the emission comes from the
S half of the ring which is in the near side of the galaxy.
Assuming that the ring is in the galactic plane, this kinematics is
that of a ring expanding radially at an average velocity of
$\approx$\,100\,km\,s$^{-1}$.
This conclusion can also be reached via the analysis of the {\it
p-v} diagrams (Sec.\,\ref{pv_H2}), in which the NE side of the ring
shows mostly redshifts, while the SW side shows blueshifts.
For an orientation of the kinematic line of nodes of $80^\circ$,
these results indicate expansion of ring in the plane.
The ring has been also observed in CO millimeter emission lines by
\citet{krips2011} who also concluded that it is expanding.

But in the {\it p-v} diagrams we have observed something else: a
velocity gradient across the ring, in which velocities of up to
$200$~km~s$^{-1}$ are observed in the inner border of the ring
(which are positive in the far side and negative in the near side)
and velocities close to zero in the outer border of the ring.
This result indicates that the expansion in the ring has negative
acceleration.
This deceleration, showing that the outer border of the ring is not
expanding anymore allows to reconcile our previous finding in
\citet{storchi12}, that the H$_2$ ring is also the site of a young
(age $\approx$ 10\,Myr) stellar population, with the ring
kinematics.
If the ring were expanding at constant velocity of
$\sim$\,100~km~s$^{-1}$, as previously thought, the 100\,pc ring
would have moved away by much more than 100\,pc in 10$^7$\,yr, and
there should be no association anymore between the H$_2$ ring and
star formation originated in this same gas.

Hydrodynamic jet simulations by \citet{gaibler12} support the
formation of rings of young stellar population in the galaxy disc as
a result of shock waves created during the early phases of the jet
activity when the jet is still contained inside the galactic disc.
They predict that these rings should have a radius of $\sim$ 100~pc.
The ring is formed by the expansion of gas pushed by the shock waves
produced by the jet which eventually stop and pile up leading to the
formation of the young stars.
The H$_2$ ring we detected in correspondence to the ring of young
stellar population \citep{storchi12} has the right size and
kinematics predicted by the model of \citet{gaibler12}.
Only that the ring is not centered at the nucleus, as predicted by
the model.
One possibility to explain this is the presence of inhomogeneities
in the gas density and distribution around the nucleus that could
lead to the off-centered appearance after the expansion.

\citet{muller09} modeled the H$_2$ kinematics in the inner tens of
pc's and proposed that there are two streams of molecular gas
towards the nucleus (indicated by NT and ST in Fig.
\ref{fig-channel-map-h2}).
In their model the ST is in front of the central source obscuring it
and flowing into it.
In their measurements the NT has an almost constant velocity of
$-25$~km~s$^{-1}$ and the ST shows an increase in velocity from
$30$~km~s$^{-1}$ at $r = 0\farcs4$ (distance from the nucleus) to
$90$~km~s$^{-1}$ at $r = 0\farcs04$.
In our data the ST has redshift velocities spanning the range $0$ to
$200$~km~s$^{-1}$ as can be seen in Fig. \ref{fig-pv-h2} in slits
$0.02$ and $-0.16$ arcsec.
The NT has blueshift velocities in the range $0$ to
$-300$~km~s$^{-1}$ as seen in the slits $0.2$, $0.38$ and $0.56$
arcsec.
These velocities seem incompatible with those of \citet{muller09}.

The analysis of our data led us to propose an alternative scenario
for the dynamics of the H$_2$ NT and ST structures.
The NT gas emission is projected in the plane of the sky
co-spatially to the brightest Br$\gamma$ emitting structure which is
seen in the channels $-47$~km~s$^{-1}$ to $-313$~km~s$^{-1}$ and is
interpreted as gas in outflow at high galactic latitude.
In order to survive, the molecular gas in the NT must be shielded
from the ionizing radiation.
We propose that the ionized gas pushes molecular gas which is thus
also observed in outflow.
The same scenario is also proposed for the ST with the difference
that this component surrounds the ionized gas in redshift which
consequently produces H$_2$ velocities in the redshift domain.
Additionally we note that in our data (as well as in those from
\citet{muller09}) the NT is brighter than the ST suggesting that the
NT is in front of the galactic disc and the ST is behind the
galactic disc.

\begin{figure*}
  \includegraphics[width=\textwidth]{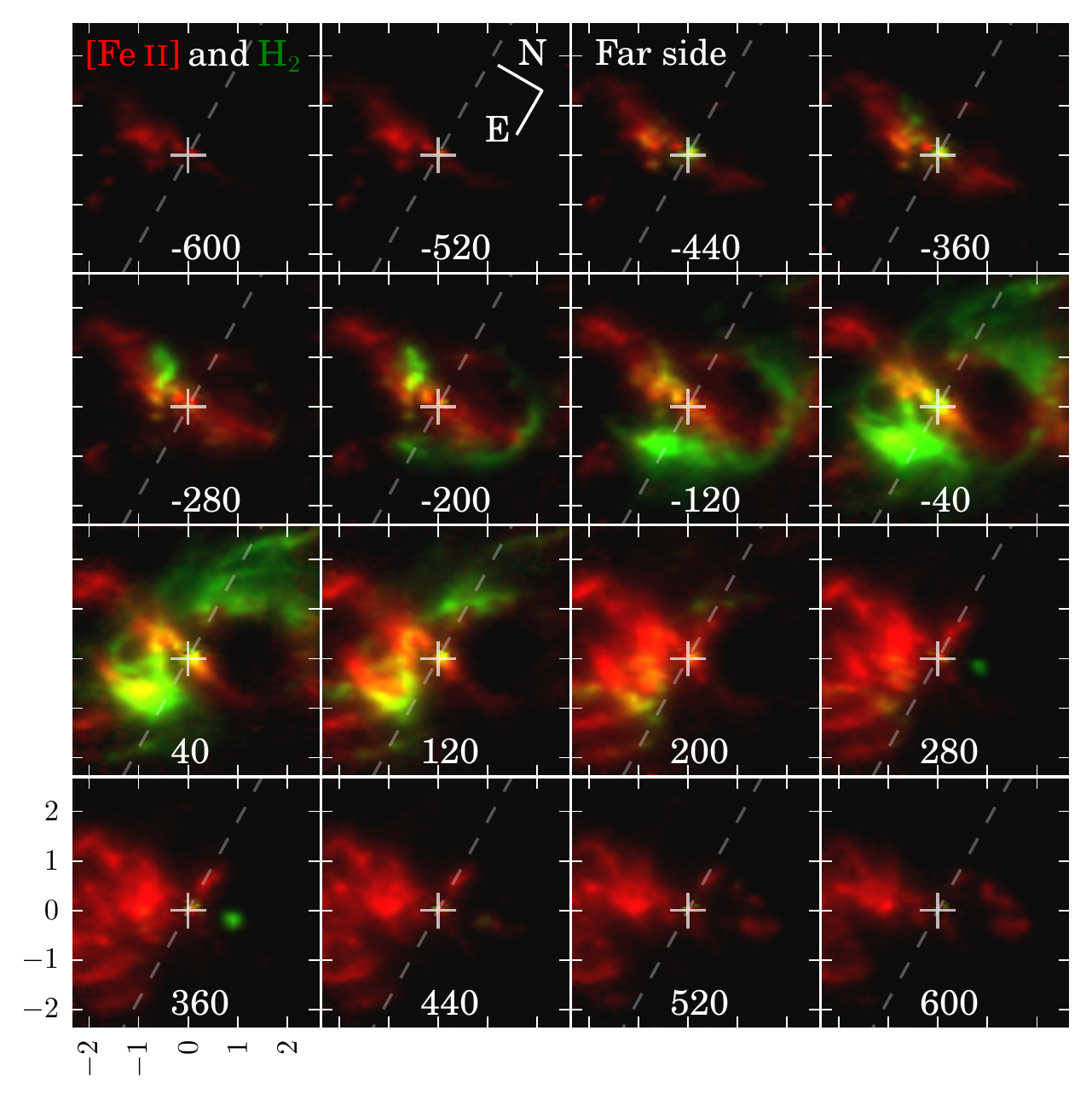}
  \caption{
  Comparison between the [Fe\,{\sc ii}] (red) and H$_2$ (green)
  kinematics.
  The dashed line shows the line of nodes.
  The center cross marks the position of the nucleus.
  The other labels are as in previous figures.
  }
  \label{fig--h2+feII}
\end{figure*}

In above scenario the H$_2$ emission is not restricted to the galaxy
plane.
We show in Figure \ref{fig--h2+feII} a comparison between the H$_2$
(green) and [Fe\,{\sc ii}] (red) kinematics via their channel maps.
In the channels $-360$~km~s$^{-1}$ and $-200$~km~s$^{-1}$ there is a
H$_2$ structure to the N which does not seem to be in the galaxy
plane.
This structure seems to connect to the NT at the channels
$-120$~km~s$^{-1}$ and $-40$~km~s$^{-1}$ supporting that the NT
originates in gas that is not in the plane.
In fact in the channel $-40$~km~s$^{-1}$ the yellow color indicate
that the NT is superimposed to the N wall of the [Fe\,{\sc ii}]
outflow.
Other superposition regions (yellow color) between H$_2$ and
[Fe\,{\sc ii}] appear in the BS area in channels $40$~km~s$^{-1}$
and $120$~km~s$^{-1}$ which suggest that the enhanced H$_2$ emission
there is related to the outflow observed in [Fe\,{\sc ii}].
Another high velocity structure which does not seem to be in the
plane is the HVC seen from channels $280$~km~s$^{-1}$ and
$440$~km~s$^{-1}$.
This structure seems to be associated to the R1 which we interpret
as part of the back wall of the SW side of the outflow (see
\ref{subsec-ionized-gas}).

\section{Outflow models}

The [Fe\,{\sc ii}] integrated flux map presents an emitting
structure with the shape similar to an hourglass with axis along the
direction NE to SW.
The NE side is brighter and more extended than the SW side.

\citet{das06}, using {\it HST/STIS} long slit spectroscopic [O\,{\sc
iii}] data proposed a kinematic model in which a hollow conical
outflow first accelerates from zero to 2000~km~s$^{-1}$ then
decelerates to zero.
The radial distance from the nucleus where the velocity reaches its
maximum value is 140~pc and where the velocity returns to zero is
400~pc.
The inner and outer cone aperture angles are, respectively,
20$^\circ$ and 40$^\circ$.
The PA and the inclination of the cone axis relative to the plane of
the sky are 30$^\circ$ and 5$^\circ$, respectively.
We have implemented this model and compared its predictions to our
[Fe\,{\sc ii}] kinematics.
The deviations from the model predictions to the [Fe\,{\sc ii}]
kinematics led us to try distinct models but having it as a starting
point.

\subsection{Conical outflow model}

\begin{figure}
  \includegraphics[width=0.5\textwidth]{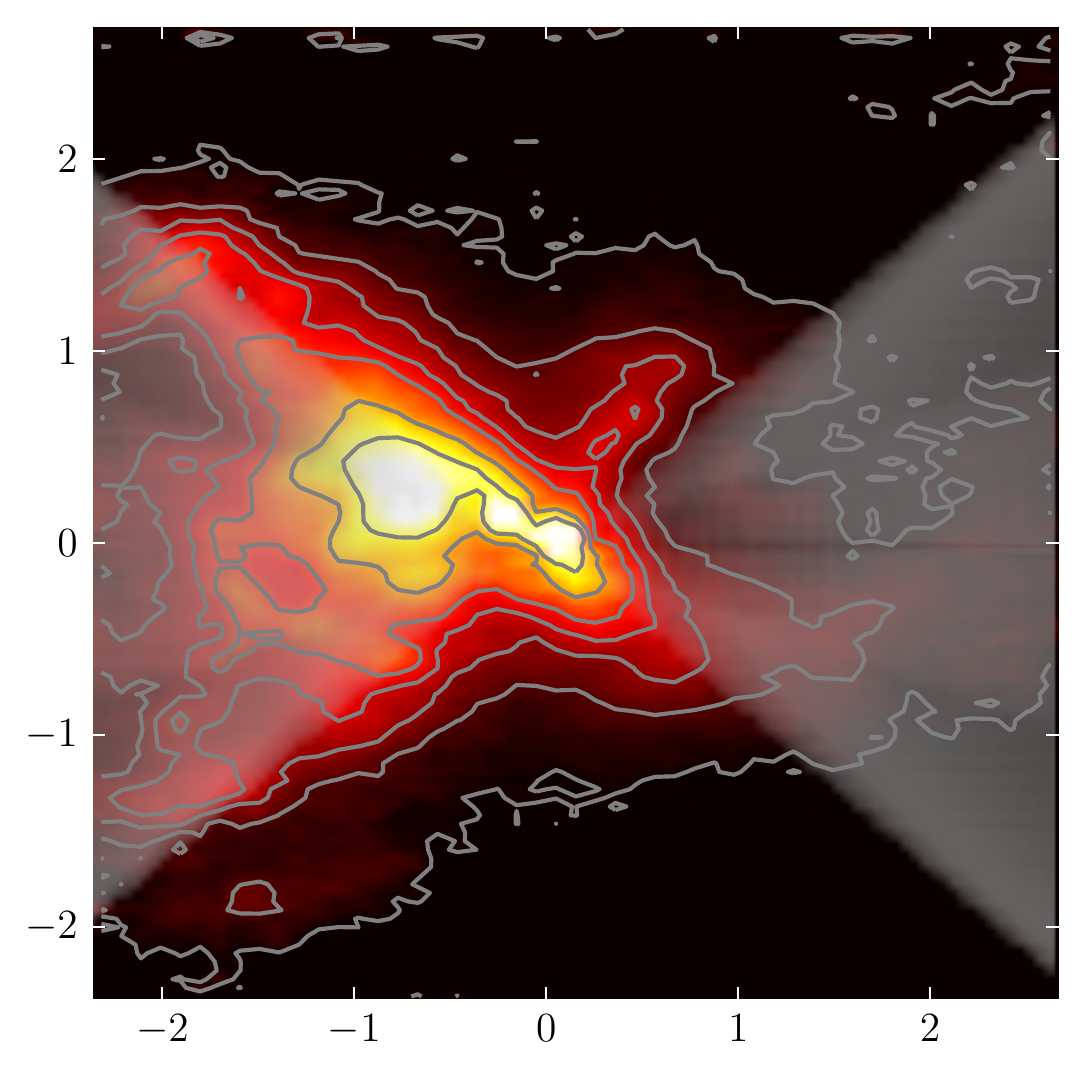}
  \caption{
  [Fe\,{\sc ii}] integrated flux map (hot colours) and hollow
  conical outflow model (grey levels).
  }
  \label{fig-flux-mod-24}
\end{figure}

In Figure \ref{fig-flux-mod-24} we present the [Fe\,{\sc ii}]
integrated flux map compared to the flux map obtained for a hollow
conical outflow model similar to that of \citet{das06}.
The model shows rough correspondence to the data but the opening
angle of the cone is smaller than the opening angle we see in the
data.
We also note that the flux map shows a change of the opening angle
with distance from the nucleus.

The conical model shown here is a modified version of the
\citet{das06} model in which the maximum velocity is lower, namely
1000~km~s$^{-1}$ instead of 2000~km~s$^{-1}$ as it provided a better
fit to our data.
In our data we do not see a decrease in velocity within our
field-of-view therefore we further modified the model to keep the
velocity at the maximum value up to the borders of the field.
Besides these changes in velocity domain, the other parameters of
their cone model were kept the same.

Summarizing, the model consists of an accelerated hollow conical
outflow with internal and external half-opening angles
$\theta_\mathrm{inner} = 20^\circ$ and $\theta_\mathrm{outer} =
40^\circ$ respectively, inclination and position angle of the bicone
axis, respectively, $i_\mathrm{axis} = 5^\circ$ (northeast is
closer) and PA$_\mathrm{axis} = 30^\circ$, the radius where the
maximum velocity occurs $r_t = 80$~pc and the maximum velocity
$v_\mathrm{max} = 1000$~km~s$^{-1}$.

The comparison between the [Fe\,{\sc ii}] and the model is shown in
Fig. 1 of the Appendix (available online as a supplementary
material).

Comparison of the model with the data shows good correspondence in
low velocity redshift channels to the NE.
The highest blueshift channels seem to show poor agreement but this
may be due to the fact that the actual outflow is patchy, in
particular in the regions corresponding to the front wall of the NE
side of the outflow.
To the SW, most of the outflow is either also very patchy or hidden
behind the galactic disc.

The {\it p-v} diagrams of the hollow conical outflow model are
compared to the [Fe\,{\sc ii}] data in Fig. 2 of the Appendix
(available online as a supplementary material).
The model shows good agreement with the data mainly in the redshift
side of the NE emission (bottom right of the diagrams).
The {\it p-v} structures to the SW side show poor agreement but, as
pointed out above, this is due to the patchy nature of the outflow
and because the SW side is partially hidden by the galactic disc.

\subsection{Spherical outflow model}

In order to address the morphological differences between the data
and the conical model and to try to meet the broader shape of the
outflow at its base we next implemented a hollow spherical outflow
model with constant velocity.
The model consists of two thick spherical cavities with centres
along the outflow axis separated by a distance $r_1 + r_2$ (where
$r_1$ is the radius of the sphere to the NE and $r_2$ is the radius
to the SW) so that the spheres touch each other at the nucleus.
The outflow is truncated at the distance of one radius from the
nucleus to both sides so that only the hemispheres close to the
nucleus compose the model.
The thickness of the spherical walls was set to $0.2\,r$.

\begin{figure}
  \includegraphics[width=0.5\textwidth]{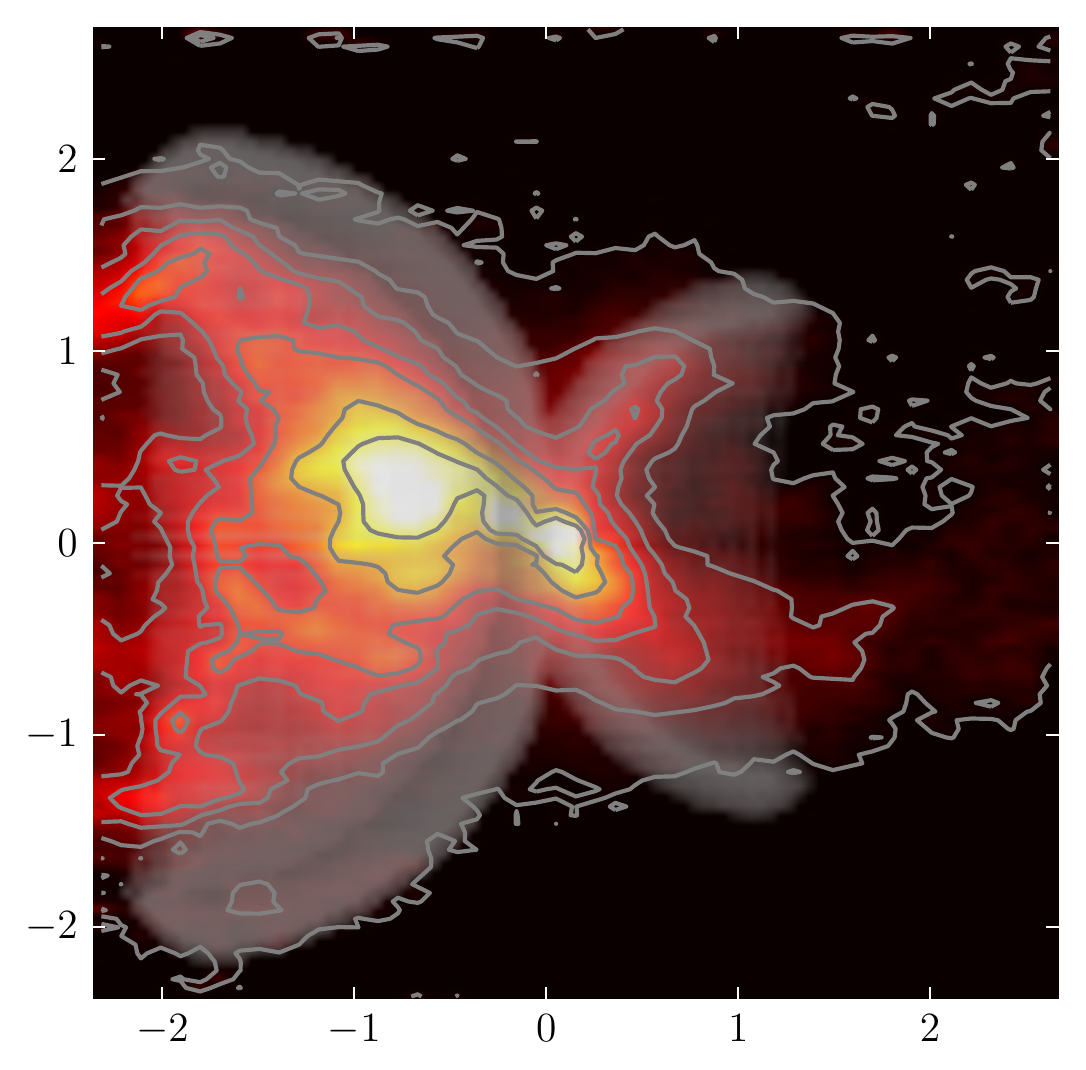}
  \caption{[Fe\,{\sc ii}] integrated flux map (hot colours) and
  hollow spherical outflow model (grey levels).
  }
  \label{fig-flux-mod-15}
\end{figure}

The flux map of the data and model is shown in Figure
\ref{fig-flux-mod-15}.
The spherical model has radii $r_1 =140$~pc and $r_2 =93$~pc,
maximum velocity $v_\mathrm{max} = 800$~km~s$^{-1}$, inclination and
position angle of the symmetry axis $i_\mathrm{axis} = 5^\circ$
(northeast is closer) and PA$_\mathrm{axis} = 30^\circ$,
respectively.

The spherical model did not fit well the data for any combination of
the parameter values.
This can be seen in Fig. 3 of the Appendix (available online as a
supplementary material).
The {\it p-v} diagrams of the hollow spherical outflow model are
presented in Fig. 4 (only for a few slits) of the Appendix
(available online as a supplementary material).
We conclude that this model is not a good fit to the data.

\subsection{Lemniscate outflow model}

\begin{figure}
  \includegraphics[width=\columnwidth]{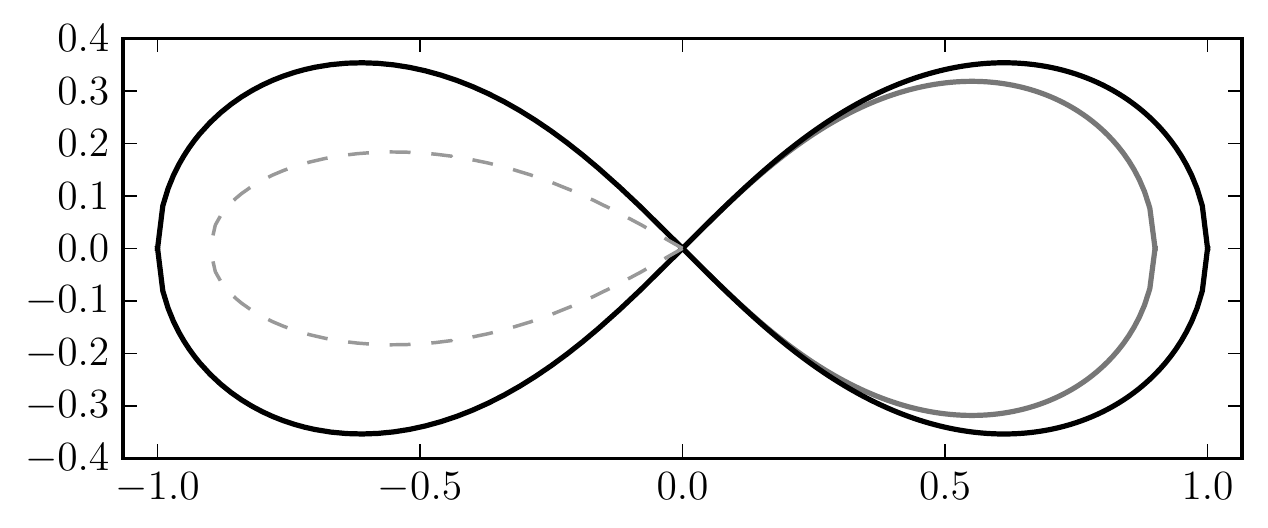}
  \caption{
  The solid black line is a regular lemniscate.
  The solid gray line (drawn to the right only) is a regular
  lemniscate with scale parameter $a$ reduced by $10$ per cent.
  The dashed gray line (seen only at left) is a modified lemniscate
  with opening angle equal to $60^\circ$ and scale parameter $a$
  reduced by $10$ per cent.
  }
  \label{fig-lemniscate}
\end{figure}

With the spherical model not being able to reproduce the data we
used a shape which is ``intermediate'' between a conical and a
spherical shape, with conical apex (at nucleus) and changing opening
angle as a function of distance to the nucleus.
We used the shape of a Lemniscate (plot in Figure
\ref{fig-lemniscate}), defined by the expression:
\begin{equation}
( x^2 + y^2 )^2 = a^2 ( x^2 - y^2 ).
\label{eq-lemniscate}
\end{equation}
The parameter $a$ is the horizontal length of the lemniscate.
Solving eq.~\ref{eq-lemniscate} for $y$ in the domain $x > 0$ we
obtain
\begin{equation}
y = \pm \frac{1}{\sqrt{2}}
    (\sqrt{8 x^2 a^2 + a^4} - 2 x^2 - a^2)^\frac{1}{2}.
\label{eq-lemniscate-y}
\end{equation}
This equation draws the right side of the lemniscate (solid black
line in Fig. \ref{fig-lemniscate}) and it is straightforward to
verify that its opening angle at the origin is always $90^\circ$.
In order to construct a physical model with this shape we must be
able to define a wall with finite width.
A first approach to accomplish this is to define internal and
external limits using lemniscates with different sizes and
populating the model in between.
The problem in using lemniscates with different sizes is that they
always have $90^\circ$ opening angles at the origin and, therefore,
close to the nucleus the wall would have near zero width (see Fig.
\ref{fig-lemniscate}) which, in turn, gives a region with no data.
We, then, modified eq. \ref{eq-lemniscate-y} to allow to set the
desired opening angle multiplying the whole equation by
$\tan{(\frac{\alpha}{2})}$:
\begin{equation}
y_2 = \tan{(\frac{\alpha}{2})}\ y,
\label{eq-lemniscate-y-modified}
\end{equation}
where $\alpha$ is the desired opening angle and $y_2$ is the new
function for the lemniscate.
The new generated geometry is that of a modified lemniscate which
has improved properties relative to the conical and spherical
models: an adjustable opening angle at the apex which decreases as a
function of distance to the nucleus.

In the velocity domain, this model consists of an acceleration
region close to the nucleus, as in the conical model, where the
velocity increases from zero at the nucleus to the maximum value
$v_\mathrm{max}$ at $r_t$ followed by a region of constant velocity
where the velocity is kept the same up to the limits of the field.

The adopted parameters and their values are: the size of the
internal and external lemniscates which are, respectively
$a_\mathrm{inner} = 405$~pc and $a_\mathrm{outer} = 450$~pc, the
internal and external half-opening angles $\theta_\mathrm{inner} =
35^\circ$ and $\theta_\mathrm{outer} = 55^\circ$ respectively, the
inclination of the lemniscate axis $i_\mathrm{axis} = 5^\circ$
(northeast is closer), the position angle of the lemniscate axis
PA$_\mathrm{axis} = 30^\circ$, the radius where the maximum velocity
occurs $r_t = 80$~pc and the maximum velocity $v_\mathrm{max} =
1000$~km~s$^{-1}$.

\begin{figure}
  \includegraphics[width=\columnwidth]{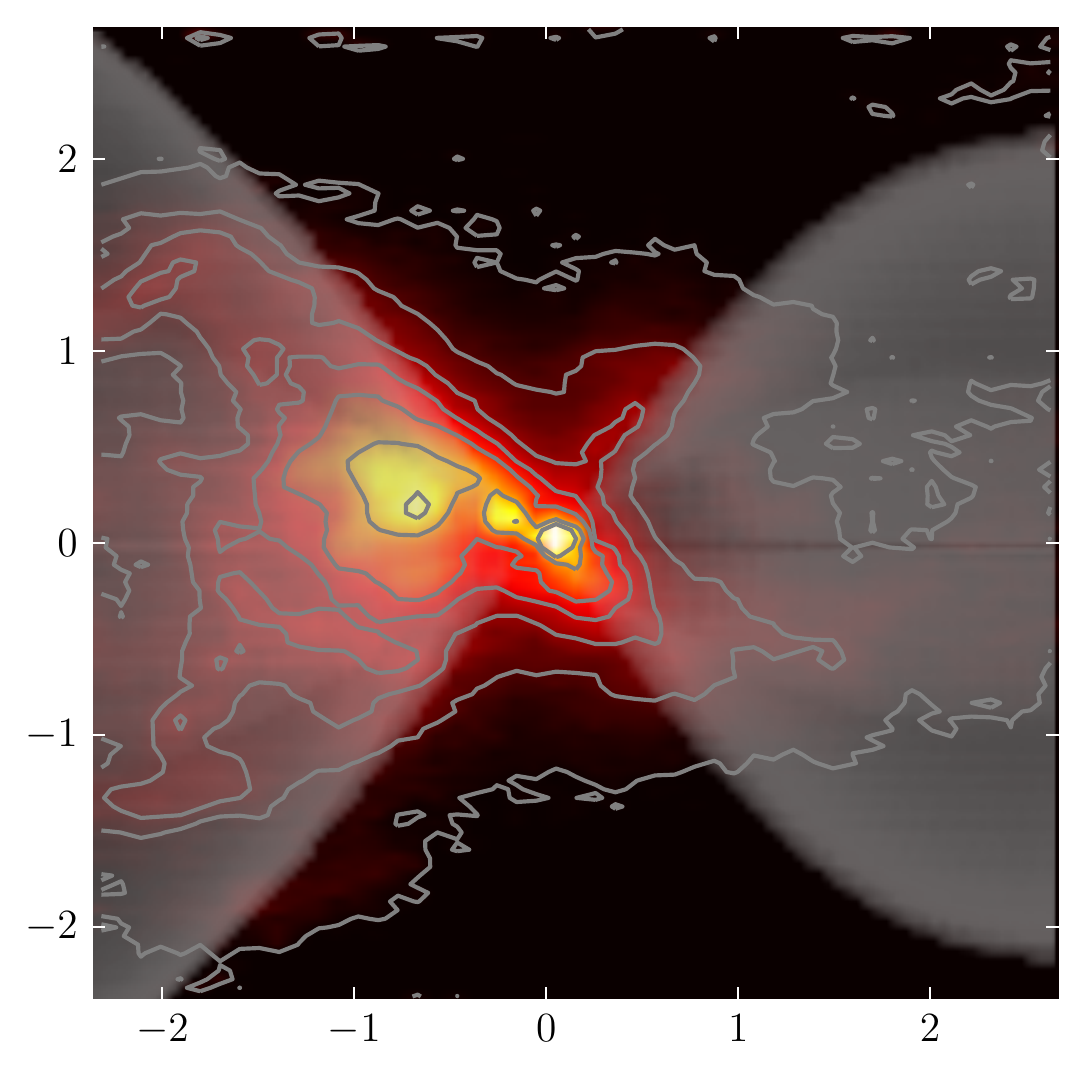}
  \caption{
  [Fe\,{\sc ii}] integrated flux map (hot colours) and hollow
  lemniscate outflow model (grey levels).
  }
  \label{fig-flux-mod-27}
\end{figure}

The integrated flux map of the lemniscate model is shown in Figure
\ref{fig-flux-mod-27} overlaid to the [Fe\,{\sc ii}] data.
Analysing only the morphological and geometrical aspects, the
lemniscate model shows some improvement over the conical model.

\begin{figure*}
  %\vspace*{174pt}
  \includegraphics[width=\textwidth]{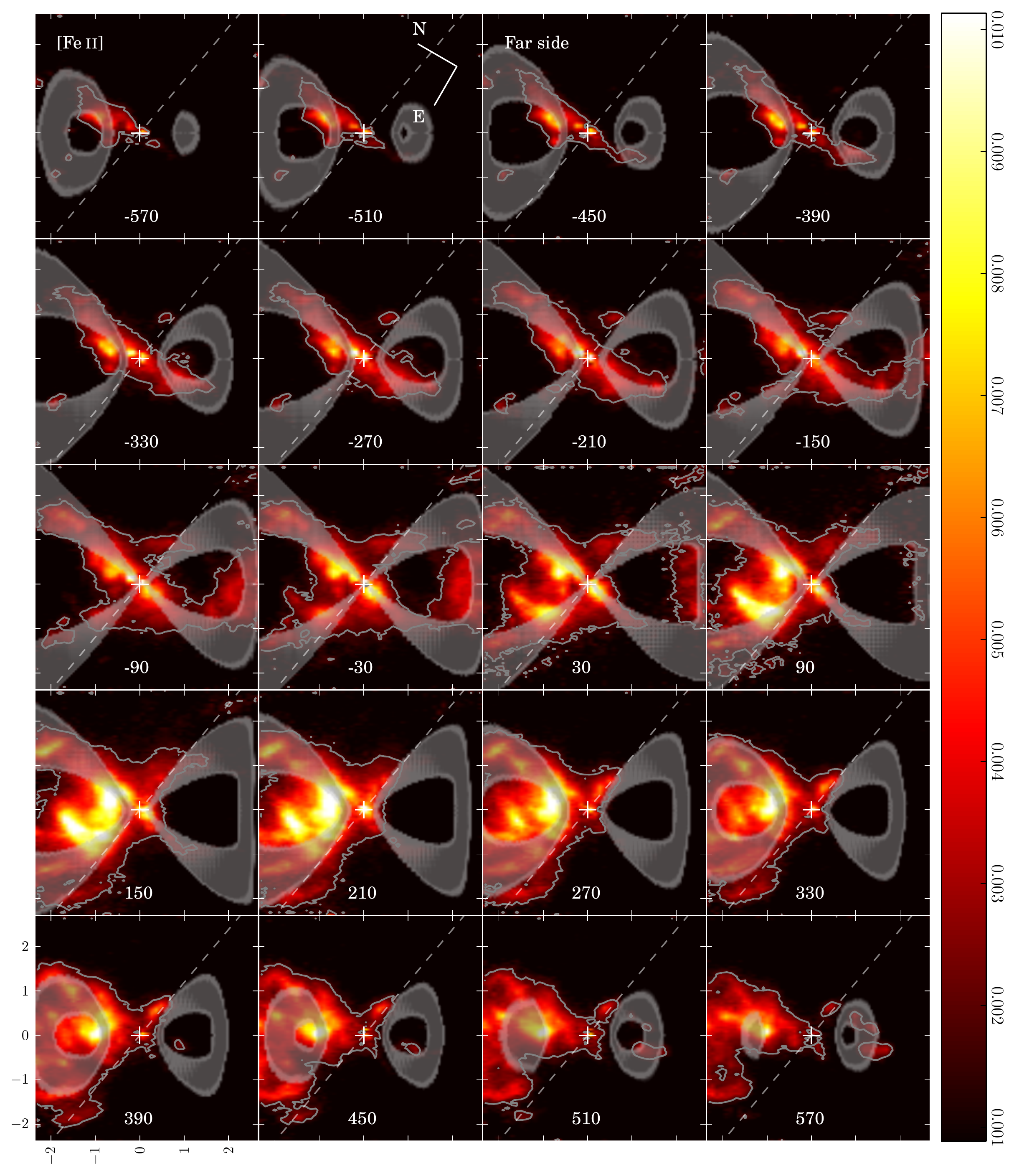}
  \caption{
  Channel maps of the accelerated lemniscate outflow model for the
  [Fe\,{\sc ii}] line (grey levels) superimposed to data (hot colours).
  The data was interpolated in wavelength to the same velocity range
  as the model.
  The central velocity of each channel is displayed in km~s$^{-1}$
  at the bottom of the corresponding panel.
  The dashed line is the adopted major axis of the galaxy.
  The tick labels are given in arcsec in the lower left panel.
  The orientation and adopted far side are indicated in the upper
  panels.
  }
  \label{fig-channel-mod-27}
\end{figure*}

In Figure \ref{fig-channel-mod-27} we present the channel maps for
the data (hot colours) and the hollow lemniscate model (grey levels)
superimposed.
For most of the channels, there is a better agreement of the model
with the data than in the previous two models.
The exceptions are the highest redshift channels ($v >
510$~km~s$^{-1}$), but, as we already discussed, we have to consider
that the outflow is patchy and that most of the SW outflow is hidden
behind the galaxy plane.
Taking this fact into account, the match between the data and model
is very good.

\begin{figure*}
  %\vspace*{174pt}
  \includegraphics[width=\textwidth]{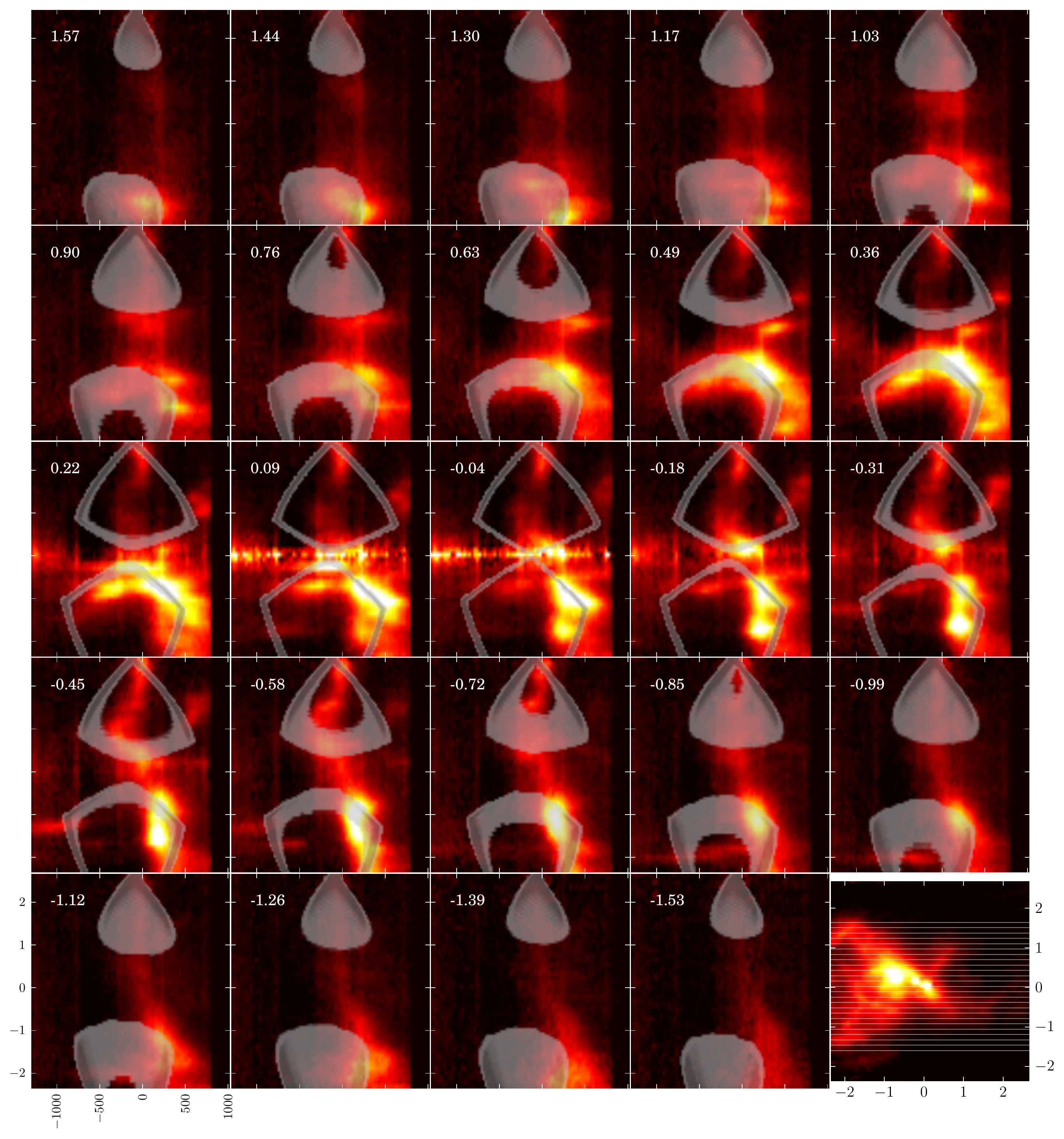}
  \caption{
  Position-velocity diagrams of the accelerated lemniscate outflow
  model for the [Fe\,{\sc ii}] line (grey levels) superimposed to
  data (hot colours).
  In the lower right panel of the figure we draw the integrated
  flux map (hot colours) and the lines corresponding to the limits
  of the adopted slits.
  The vertical displacement from the nucleus to the center of each
  slit is displayed in the upper left corner of each panel, in
  arcsec.
  }
  \label{fig-pv-mod-27}
\end{figure*}

The {\it p-v} diagrams of the hollow lemniscate outflow model are
compared to the data in Figure \ref{fig-pv-mod-27}.
There is good agreement to the data, mainly in the redshift side of
the NE emission (bottom right of the diagrams) Again, the patchy
nature of the outflow and the fact that the SW side of the outflow
is behind the galactic disc must be taken into account to explain
the discrepancies.

\subsection{Comparison of the models}

Given the distinct morphology of the [Fe\,{\sc ii}] as compared to
that of the [O\,{\sc iii}] emitting gas and to that of the conical
model, it is remarkable that the conical model previously proposed
by \citet{das06} -- based on long-slit spectroscopy -- gives such a
good reproduction of our [Fe\,{\sc ii}] data, although we have tuned
a bit the parameters.
The conical model, however gives a poorer reproduction of the
outflowing gas morphology than the lemniscate model as seen in the
integrated flux map (Fig. \ref{fig-flux-mod-24}) and in the channel
maps (Figure 1 of the Appendix, available online as a supplementary
material).

The spherical model fits poorly the data and results in no
improvement over the conical model.

The lemniscate model gives a better fit to the flux distribution,
but the improvement in the kinematics relative to the conical model
is modest: in the {\it p-v} diagrams, the lemniscate model reproduces
better the data for the slits that are farthest away from the
centre.
In the slits closest to the nucleus, there is a ``knee" in the
positive velocities at $\approx$\,1\farcs2 from the nucleus that is
also well reproduced by the lemniscate model.
It may be argued that this model shows a narrower range of
velocities than the data, particularly noticeable in the central
slits of the {\it p-v} diagrams, but this can be understood as due
to the fact that we did not include a velocity dispersion across the
walls of the lemniscate.
In the conical model, such velocity dispersion arises naturally
because of the radial symmetry of the model which leads to a varying
orientation of the velocity vectors across the walls of the cone,
while, in the same region of the lemniscate model the velocity
vectors are parallel.

We would like to point out that, even though the geometric
distribution of the gas seems to be better reproduced by the
lemniscate model, there is no significant difference between the
kinematics of the conical and lemniscate models, both providing
similar reproduction of the data.
The conical model and the lemniscate model give equally good (or
bad!) reproduction of the kinematic data.
This is due to the fact that our modeling did not include all the
effects present in the data, only a simple geometry and kinematics
for the outflow.
No velocity component due to gas in the disc was considered in the
models, even though they seem to be present in the data.
We have also not considered the patchy nature of the outflow, which
is obvious in the channel maps, neither the enhanced emission in the
redshift channel maps to the NE, as observed in the data.
As we have already pointed out, these latter features are due to the
interaction between an accretion disc outflow and the circumnuclear
ISM of the galaxy.
Due to the orientation of the outflow relative to the galaxy plane,
the back wall of the NE part of the outflow is tangent to the galaxy
plane, encountering much more gas on its way out than the front wall
which leaves the plane at a larger angle.

The absence of the effects discussed above in the modelling
precludes a quantitative discrimination of the best model to fit the
data.
As we have stated above we can only say that the lemniscate model
gives a somewhat better reproduction of the line flux distributions.
An additional discriminator between the models could be obtained if
the outflow could be probed at larger distances from the nucleus
than we have observed, as the models predict distinct morphologies
and behaviors in that regions.

The outflow we are seeing on scales of tens to hundreds of parsecs
originates in an accretion disc outflow which occurs on scales 2--3
orders of magnitudes smaller.
We are thus observing the interaction of an AGN outflow with the
circumnuclear gas at the galaxy.
Thus it is remarkable that the observed geometry is so similar to
that proposed in accretion disc wind models, such as those of
\citet{proga04} and \citet{proga00}.
The fact that the outflow is hollow supports also an origin in a
finite range of radii in the accretion disc as proposed by
\citet{elvis00}.
The only difference between the AGN outflow models and the data is
the fact that the opening angle of the observed outflow decreases
with distance from the nucleus, what seems not to be predicted by
the models and may be due to the interaction of the AGN outflow and
the circumnuclear gas of the galaxy.
It is interesting to note that this feature is observed also in some
planetary nebulae (e.g. NGC\,6302 and NGC\,6537).

\section{Mass outflow rate}

In order to quantify the mass outflow rate, consider a cross section
of the lemniscate perpendicular to the model axis at a distance of
1~arcsec from the nucleus.
This cross section has the geometry of a ring with area $A$.
The mass outflow rate is
\begin{equation}
\dot{M}_{out} = m_{p}\,n_{e}\,v_{\mathrm{perp}}\,f\,A,
\label{mass_outflow}
\end{equation}
where $m_p$ is the proton mass, $n_e$ the electron density,
$v_{\mathrm{perp}}$ is the velocity perpendicular to the cross
section and $f$ is the filling factor.

For the velocity $v_{\mathrm{perp}}$ we adopt the velocity given by
the model at the cross section, $900$~km~s$^{-1}$, corrected for its
orientation relative to the direction perpendicular to the cross
section at this location, $40.8^\circ$.
The result is $v_{\mathrm{perp}} = 900\,\cos{(40.8^\circ)} \simeq
680$~km~s$^{-1}$.

The filling factor $f$ can be obtained from

\begin{equation}
L_{\mathrm{H\beta}} = 4\,\pi\,j_{\mathrm{H\beta}}(T)\,V\,f,
\label{filling-factor}
\end{equation}
\noindent
where $L_{\mathrm{H\beta}}$ is the H$\beta$ luminosity emitted by a
volume $V$ and
$4\,\pi\,j_{\mathrm{H\beta}}(T) =
(4\,\pi\,j_{\mathrm{H\beta}}(T)/n_e\,n_p) \times n_e\,n_p =
1.23\times10^{-25} \times 10^4 =
1.23\times10^{-21}$~erg~cm$^{-3}$~s$^{-1}$ \citep{osterbrock06}
assuming $n_e = n_p = 10^2$~cm$^{-3}$ and $T = 10000$~K (Case B). 
Although the temperature and density of the gas are conservative
guesses, the value of $4\,\pi\,j_{\mathrm{H\beta}}(T)$ for a wide
set of possible temperatures and densities ranges from 0.66 to $2.3
\times10^{-21}$~erg~cm$^{-3}$~s$^{-1}$.
$L_{\mathrm{H\beta}}$ is determined from the flux
$F_{\mathrm{Br\gamma}}$ using

\begin{equation}
L_{\mathrm{H\beta}} = 4\,\pi\,D^2\,F_{\mathrm{H\beta}} =
4\,\pi\,D^2\,F_{\mathrm{Br\gamma}}\ /\ \frac{j_{\mathrm{Br\gamma}}}{j_{\mathrm{H\beta}}},
\label{eq_lum_hbeta_2}
\end{equation}
\noindent
where $D$ is the distance to the galaxy and $j_{\mathrm{Br\gamma}}$
and $j_{\mathrm{H\beta}}$ are emission coefficients for Br$\gamma$
and H$\beta$, respectively.

The flux $F_{\mathrm{Br\gamma}}$ was obtained integrating the
Br$\gamma$ flux map over a region with the geometry of a right
triangle to the N-NE of the nucleus such that the nucleus is at one
of the vertices with angle $50^\circ$.
This is the angle between the outflow axis (one of the triangle
sides) and the wall of the cone model (the triangle hypotenuse).
The other triangle side is placed orthogonal to the axis of the
outflow at 1~arcsec from the nucleus.
This is the region where the Br$\gamma$ is brightest.
The integrated flux over this triangular region is $0.534 \times
10^{-15}$~erg~cm$^{-2}$~s$^{-1}$.
The triangular region over which the flux was integrated corresponds
to the projection, in the plane of the sky, of one half of the
volume of a cone with the same cross section area of the lemniscate
model.

Now dividing equation \ref{filling-factor} by \ref{eq_lum_hbeta_2},
solving for $f\, V$ and using that the volume $V$ in equation
\ref{filling-factor} must be the difference between the volumes of
the inner and outer half cones $V = V_o - V_i =
\frac{1}{2}\frac{h}{3}\, A_o - \frac{1}{2} \frac{h}{3}\, A_i =
\frac{h}{6} (A_o - A_i) = \frac{h}{6}\, A$ we obtain $f\, A$:

\begin{equation}
f\,A = \frac{D^2\ F_{\mathrm{Br\gamma}}}{h/6\, j_{\mathrm{Br\gamma}}}.
\label{eq_fA}
\end{equation}
Although we are calculating the outflow for a lemniscate geometry
and using a conical volume in equations \ref{filling-factor} and
\ref{eq_lum_hbeta_2} (the volume is implicit in the integrated
flux), this is no mistake as the volume in both equations was the
same and its dependence cancel out giving place to a dependence in
the area $A$, which is the same as that of the lemniscate model. 
It is also assumed that the filling factor $f$ is the same for the
entire cross section.
Inserting \ref{eq_fA} in \ref{mass_outflow} we obtain
\begin{equation}
\dot{M}_{out} = m_{p}\,n_{e}\,v_{\mathrm{perp}}\,\frac{D^2\
F_{\mathrm{Br\gamma}}}{h/6\, j_{\mathrm{Br\gamma}}}
\label{eq_mass_outflow_simp}
\end{equation}
which does not depend explicitly on geometric details of the
outflow.

Now using equation \ref{eq_mass_outflow_simp} we obtain a mass
outflow rate of $\dot{M}_{out} = 1.9~M_\odot~\mathrm{yr}^{-1}$.
This value is in agreement with those typical of other active
galaxies in \citet{veilleux05}, which range from 0.1 to
$10~M_\odot~\mathrm{yr}^{-1}$ and also with that obtained by
\citet{mrk1157}, of $6~M_\odot~\mathrm{yr}^{-1}$, for Mrk\,1157 as
well as with the revised value for Mrk\,1066 \citep{mrk1066}.
As pointed out also in the above studies, such mass outflow rates
are much larger than the AGN accretion rate, indicating that,
although the origin of the outflow is the AGN, the observed outflow
is actually gas from the circumnuclear region being pushed by the
AGN outflow (``mass-loaded outflow'').

The $\dot{M}_{out}$ value depends mainly on the uncertainties of
$v_\mathrm{perp}$, $D$, $F_{\mathrm{Br\gamma}}$ and
$j_{\mathrm{Br\gamma}}$ and on the assumptions of the modeling.
From these parameters the less constrained are the
$j_{\mathrm{Br\gamma}}$ and $D$.
As mentioned above $j_{\mathrm{Br\gamma}}$ can span a range from
0.66 to $2.3 \times 10^{-21}$~erg~cm$^{-3}$~s$^{-1}$ which
corresponds to about $\pm 100\%$ uncertainty.
The distance to NGC\,1068 $D$ is also very uncertain as in the
literature we find values ranging from 10 to 30~Mpc that correspond
to an uncertainty of about $^{+100\%}_{-30\%}$.
Taking into account these uncertainties, the mass outflow rate is in
the range 1 to $4~M_\odot~\mathrm{yr}^{-1}$

We can use the above mass outflow rate to estimate the mechanical
power of the outflow using the same method as in \citet{barbosa09}:
\begin{equation}
P = \frac{1}{2}\,\dot{M}_{out}\,v^{2}_\mathrm{perp} .
\end{equation}
This is a lower limit as this formula does not account for the power
associated to the velocity dispersion of the gas, however, this
additional term can be neglected as the velocity dispersion is much
smaller than the velocity of the outflow.
Using the above $v_\mathrm{perp}$, we obtain $P \cong 2.8\times
10^{41}$~erg~s$^{-1}$, which is in good agreement with those
obtained for Seyfert galaxies and compact radio sources
\citep{morganti05} and with that found in \citet{mrk1157}, of $P
\cong 5.7\times10^{41}$\,erg\,s$^{-1}$ for Mrk\,1157. 
The uncertainties for the mechanical power are, in our case, the
same as those for $\dot{M}_{out}$ and allow for $P$ values in the
range 1.5 to $5.6\times 10^{41}$~erg~s$^{-1}$.
Adopting a bolometric luminosity of $L_{Bol} \cong 3.6 \times
10^{44}$~ergs~s$^{-1}$ from \citet{pier1994} (the authors argue that
the $L_{Bol}$ is probably within a factor of a few of the
published value) for the nuclear source, the mechanical power $P$ is 
within a factor of a few of $0.08\%~L_{Bol}$.
The typical power adopted in AGN feedback models to significantly
affect the galaxy evolution \citep[e.g.][]{dimatteo2005} is
$0.5\%~L_{Bol}$ and, consequently, the outflow mechanical power in
NGC\,1068 is smaller than that required to provide significant
feedback effect in the galaxy.
This conclusion depends also on the adopted $L_{Bol}$, as a 7 times
smaller value would lead to the opposite conclusion.
\citet{woo02} estimated $L_{Bol} \cong 9.55 \times10^{44}$ and
\citet{prieto10} calculated $L_{Bol} \cong 8.7 \times10^{43}$.
These values place the limits for our calculated mechanical power in
the range $0.03\%~L_{Bol}$ to $0.3\%~L_{Bol}$.
Nonetheless, the values of $L_{Bol}$ for NGC\,1068 must be taken as
conservative lower limits, as the $A_V$ of the nuclear source is
known to be high and this can lead to underestimated values of
$L_{Bol}$.
These factors lead us to argue that the mechanical power in
NGC\,1068 is smaller than that required to provide significant
feedback effect to the galaxy evolution.

\section{Summary and Conclusions}

We have mapped the inner (200\,pc) NLR kinematics of the Seyfert 2
galaxy NGC\,1068 on the basis of near-IR integral-field spectroscopy
in the emission lines [Fe\,{\sc ii}]\,$\lambda\,1.644\,\mu$m,
H$_2\,\lambda\,2.122\,\mu$m and Br$\gamma$.

We have compared the ionized and molecular gas kinematics and have
used three different models to reproduce channel maps and {\it p-v}
diagrams of the ionized gas.

Our main conclusions are:

\begin{itemize}

\item The Br$\gamma$ kinematics comprises both blueshifts and
redshifts of up to $1000$\,km\,s$^{-1}$ and its flux distribution in
the channel maps is very similar to the flux distribution in
[O\,{\sc iii}], confirming our conclusion of Paper~I that the two
originate in the same bi-polar outflow.

\item The [Fe\,{\sc ii}] kinematics cover the same bi-polar outflow
and range of velocities as observed for Br$\gamma$, but the flux
distributions in the channel maps show a broader (bowl-shaped
hourglass) bi-polar morphology than the bi-conical shape previously
observed in the [O\,{\sc iii}] flux distribution and also observed
in Br$\gamma$.
This is mainly due to the larger contribution of [Fe\,{\sc ii}]
emission in the redshifted channels.
We attribute this result to the origin of the [Fe\,{\sc ii}]
emission in a partially ionized region extending beyond the fully
ionized region probed by the Br$\gamma$ and [O\,{\sc iii}] emission,
with additional contribution from shocks between the nuclear outflow
and gas in the galaxy disk.

\item We conclude that the [Fe\,{\sc ii}] kinematics provide a
better coverage of the NLR outflow thus being a better tracer of the
outflow than both [O\,{\sc iii}] and H$^+$ emission lines, as it
extends to regions not probed by these lines.

\item The bi-polar bowl shape geometry (including the partially
ionized gas region) resembles that of many planetary nebulae and
that predicted by disc and torus wind models, supporting that such
winds are the origin of the observed outflow.

\item The H$_2$ kinematics is completely distinct from that of the
ionized gas, showing an off-centered ring-like morphology at much
lower velocities that suggest expansion in the galaxy plane, at an
average velocity of $\approx$\,100\,km\,s$^{-1}$ as supported by
previous CO mm observations.

\item A new result regarding H$_2$ is our finding of deceleration
between the inner and outer border of the ring from
$200$\,km\,s$^{-1}$ down to zero, which indicates that the expansion
is being halted.
This result has been predicted in recent hydrodynamic simulations
and can be attributed to the interaction of a nuclear jet with the
gas in the galaxy disk.
This gas is pushed radially away from the nucleus and subsequently
halted by the circumnuclear ISM.
This deceleration reconciles our previous finding of a 10\,Myr
stellar population in the ring with its kinematics.

\item Our measurements also suggest that the two H$_2$ linear
structures NT and ST (to the north and south of the nucleus) which
apparently connect the ring to the galaxy nucleus are in outflow
(contrary to the suggested inflows in previous studies).

\item We have built models testing three distinct geometries for the
outflow observed in [Fe\,{\sc ii}]: a conical, a spherical and a
lemniscate (hourglass).
Although the lemniscate model gives a better reproduction of the
integrated flux distribution, its kinematics does not provide a
significantly better reproduction of the velocity field than the
conical model of \citet{das06}; in both models, the outflow is
accelerated to a maximum velocity of 1000\,km\,s$^{-1}$ at 80\,pc
from the nucleus, decelerating to zero in the conical model and
keeping its maximum velocity constant in the lemniscate model for
distances larger than 80\,pc from the nucleus.

\item The calculated mass outflow rate along the NLR is
$1.9^{+1.9}_{-0.7}$\,M$_\odot$\,yr$^{-1}$.
Similarly to what we and others have found for other Seyfert
galaxies, this outflow rate is much larger than the mass accretion
rate to the AGN, indicating that the observed outflow is in fact gas
from the inner region of the galaxy pushed by an AGN outflow.

\item The calculated power of the outflow is
$2.8^{+2.8}_{-1.3}\times10^{41}$~erg~s$^{-1}$, which is only
$0.08\%\,L_{Bol}$, thus smaller than the $0.5\%\,L_{Bol}$
required to affect significantly the galaxy evolution.

\end{itemize}

\section*{Acknowledgments}

We thank the referee for valuable suggestions that helped to improve
the paper.
Based on observations obtained at the Gemini Observatory, which is
operated by the Association of Universities for Research in
Astronomy, Inc., under a cooperative agreement with the NSF on
behalf of the Gemini partnership: the National Science Foundation
(United States), the National Research Council (Canada), CONICYT
(Chile), the Australian Research Council (Australia), Minist\'{e}rio
da Ci\^{e}ncia, Tecnologia e Inova\c{c}\~{a}o (Brazil) and
Ministerio de Ciencia, Tecnolog\'{i}a e Innovaci\'{o}n Productiva
(Argentina).

\bsp

\label{lastpage}


\begin{thebibliography}{99}

%\bibliography{p}


%\bibitem[\protect\citeauthoryear{Aguerri, Beckman \&
%Prieto}{1998}]{ABP98} Aguerri J.A.L., Beckman J.E., Prieto M.,
%1998, AJ, 116, 2136

%\bibitem[\protect\citeauthoryear{Allington-Smith et al.}{2002}]{all02} 
%Allington-Smith J., et al., 2002, PASP, 114, 892

%\bibitem[\protect\citeauthoryear{Andrillat \& Souffrin}{1968}]{AS68}
%Andrillat Y., Souffrin S., 1968, ApL, 1, 111

%\bibitem[\protect\citeauthoryear{Arribas et al.}{1997}]{arr97}
%Arribas S., Mediavilla E., Garcia-Lorenzo B., del Burgo C., 1997,
%ApJ, 490, 227

%\bibitem[\protect\citeauthoryear{Barbosa et al.}{2006}]{barbosa06}
%Barbosa F.K.B., Storchi-Bergmann T., Cid Fernandes R., Winge C., Schmitt H.,
%2006, MNRAS, 371, 170

\bibitem[\protect\citeauthoryear{Barbosa et al.}{2009}]{barbosa09}
Barbosa, F.K.B., Storchi-Bergmann, T., Cid Fernandes, R., Winge, C., 
\& Schmitt, H., 2009, MNRAS, 396, 2

\bibitem[\protect\citeauthoryear{Cecil et al.}{2002}]{cecil02}
Cecil, G., Dopita, M. A., Groves, B., Wilson, A. S. \& Binette, L., 2002, ApJ 568, 627

%\bibitem[\protect\citeauthoryear{Barth et al.}{2002}]{bar02}
%Barth A.J., Ho, L.C., Sargent W.L.W., 2002, AJ, 124, 2607

%\bibitem[\protect\citeauthoryear{Blietz et al.}{1994}]{bli94}
%Blietz M., Cameron M., Drapatz S., Genzel R., Krabbe A., van
%der Werf P., Sternberg A., Ward M., 1994, ApJ, 421, 92

%\bibitem[\protect\citeauthoryear{Boisson et al.}{2000}]{boi00}
%Boisson C., Joly M., Moultaka J., Pelat D., Serote Roos M., 2000,
%A\&A, 357, 850

%\bibitem[\protect\citeauthoryear{Brinks et al.}{1997}]{bri97}
%Brinks, E., Skillman, E.~D., Terlevich, R.~J., \& Terlevich, E.\
%1997, Ap\&SS, 248, 23

%\bibitem[\protect\citeauthoryear{Capetti et al.}{1996}]{cap96}
%Capetti A., Axon D.J., Macchetto F., Sparks W.B., Boksenberg A.,
%1996, ApJ, 469, 554

%\bibitem[\protect\citeauthoryear{Cappellari \& Emsellem}{2004}]{cap04}
%Cappellari M., Emsellem E., 2004, PASP, 116, 138

%\bibitem[\protect\citeauthoryear{Cenarro et al.}{2001}]{cen01}
%Cenarro A.J., Cardiel N., Gorgas J., Peletier R.F., Vazdekis A., Prada F.,
%2001, MNRAS, 326, 959

%\bibitem[\protect\citeauthoryear{Cid Fernandes et al.}{2001}]{cid01}
%Cid Fernandes R., Heckman T., Schmitt H., Delgado R.M.G.,
%Storchi-Bergmann T., 2001, ApJ, 558, 81

%\bibitem[Cid Fernandes et al.(2003)]{2003ASPC..297..357C} Cid Fernandes, 
%R., Schmitt, H., Gonz{\' a}lez Delgado, R.~M., Storchi-Bergmann, T., 
%Heckman, T., \& Rodrigues Lacerda, R.\ 2003, ASP Conf.~Ser.~297: Star 
%Formation Through Time, 297, 357 
 

%\bibitem[\protect\citeauthoryear{Contini \& Viegas}{1999}]{CV99}
%Contini M., Viegas S.M., 1999, ApJ, 523, 114

%\bibitem[\protect\citeauthoryear{Crenshaw}{1986}]{cre86}
%Crenshaw D.M., 1986, ApJS, 62, 821

\bibitem[\protect\citeauthoryear{Das et al.}{2006}]{das06}
Das V., Crenshaw D.M., Kraemer S.B., Deo R.P., 2006, AJ, 132, 620

%\bibitem[\protect\citeauthoryear{Delgado et al.}{1997}]{del97}
%Delgado R.M.G., Perez E., Tadhunter C., Vilchez J.M.,
%Rodriguez-Espinosa J.M., 1997, ApJS, 108, 155

\bibitem[\protect\citeauthoryear{de Vaucouleurs, G., et al.}{1991}]{dev91}
de Vaucouleurs, G., et al. 1991, Third Reference Catalogue of Bright
Galaxies (New York: Springer)

\bibitem[\protect\citeauthoryear{di Matteo et al.}{2005}]{dimatteo2005}
Di Matteo T., Springel V., \& Hernquist L., 2005, Nature, 433, 604

%\bibitem[\protect\citeauthoryear{Devereux}{1989}]{dev89} Devereux
%N.A., 1989, ApJ, 346, 126

%\bibitem[\protect\citeauthoryear{D{\'{\i}}az et al.}{2003}]{dia03}
%D\'{\i}az R.J., Dottori H., Vera-Villamizar N., Carranza G., 2003,
%ApJ, 597, 860

%\bibitem[\protect\citeauthoryear{van Driel \& Buta}{1991}]{vDB91}
%van Driel W., Buta R.J., 1991, A\&A, 245, 7

\bibitem[\protect\citeauthoryear{Elvis}{2000}]{elvis00}
Elvis, M.\ 2000, ApJ, 545, 63

%\bibitem[\protect\citeauthoryear{Emsellem et al.}{2001}]{ems01}
%Emsellem E., Greusard D., Combes F., Friedli D., Leon S., P{\' e}contal E.,
%Wozniak H., 2001, A\&A, 368, 52

%\bibitem[\protect\citeauthoryear{Emsellem}{2004}]{ems04}
%Emsellem, The Interplay among Black Holes, Stars and ISM in Galactic Nuclei,
%Proceedings of IAU Symposium, No. 222.  Edited by T.
%Storchi-Bergmann, L.C. Ho, and Henrique R. Schmitt.  Cambridge, UK:
%Cambridge University Press, 2004., p.419-422

%\bibitem[\protect\citeauthoryear{Erwin \& Sparke}{2003}]{ES03}
%Erwin P., Sparke L.S., 2003, ApJS, 146, 299

\bibitem[\protect\citeauthoryear{Emsellem et al.}{2006}]{emsellem06}  Emsellem, E., Fathi, K., Wozniak, H., Ferruit, P., Mundell, C. G., Schinnerer, E. 2006, MNRAS, 365, 367

\bibitem[\protect\citeauthoryear{Evans et al.}{1991}]{evans1991}
Evans, I.N., Ford, H.C., Kinney, A.L., et al., 1991, ApJL, 369, L27

\bibitem[\protect\citeauthoryear{Gerssen et al.}{2006}]{gerssen06}  Gerssen, J., Allington-Smith, J., Miller, B. W., Turner, J. E. H., Walker, A., 2006, MNRAS, 365, 29

%\bibitem[\protect\citeauthoryear{Evans}{1996}]{eva96}
%Evans I.N., Koratkar A.P., Storchi-Bergmann T., Kirkpatrick H.,
%Heckman T.M., Wilson A.S., 1996, ApJS, 105, 93

%\bibitem[\protect\citeauthoryear{Falcke et al.}{1998}]{fal98}
%Falcke H., Wilson A.S., Simpson C., 1998, ApJ, 502, 199

%\bibitem[\protect\citeauthoryear{Fernandez et al.}{1999}]{fer99}
%Fernandez B.R., Holloway A.J., Meaburn J., Pedlar A., Mundell C.G.,
%1999, MNRAS, 305, 319

%\bibitem[\protect\citeauthoryear{Ferrarese \& Merritt}{2000}]{FM00}
%Ferrarese L., Merritt, D., 2000, ApJL, 539, L9

%\bibitem[\protect\citeauthoryear{Ferruit, Wilson \& Mulchaey}{2000}]{FWM00}
%Ferruit P., Wilson A.S., Mulchaey J., 2000, ApJS, 128, 139

%\bibitem[\protect\citeauthoryear{Forbes \& Ward}{1993}]{fw93}
%Forbes D.A., Ward M.J., 1993, ApJ, 416, 150

\bibitem[\protect\citeauthoryear{Gaibler et al.}{2012}]{gaibler12}
Gaibler V., Khochfar S., Krause M., \& Silk J., 2012, MNRAS, 425, 438 

%\bibitem[\protect\citeauthoryear{Ganda et al.}{2006}]{gan06}
%Ganda K., Falc{\'o}n-Barroso J., Peletier R.F., Cappellari M.,
%Emsellem E., McDermid R.M., Tim de Zeeuw P., \& Carollo C.M., 2006,
%MNRAS, 367, 46

%\bibitem[\protect\citeauthoryear{Garc{\'{\i}}a-Lorenzo et al.}{1999}]{gar99} 
%Garc{\'{\i}}a-Lorenzo B., Mediavilla E., Arribas S., 1999, ApJ, 518, 190

%\bibitem[\protect\citeauthoryear{Garc{\'{\i}}a-Rissmann et al.}{2005}]{gar05} 
%Garcia-Rissmann A., Vega L.R., Asari N.V., Cid Fernandes R., Schmitt 
%H., Gonz{\'a}lez Delgado R.M., Storchi-Bergmann T., 2005, MNRAS, 359, 765

%\bibitem[\protect\citeauthoryear{Gebhardt et al.}{2000}]{geb00}
%Gebhardt K., et al., 2000, ApJL, 539, L13

%\bibitem[\protect\citeauthoryear{George et al.}{1998}]{geo98}
%George I.M., Mushotzky R., Turner T.J., Yaqoob T., Ptak A.,
%Nandra K., Netzer H., 1998, ApJ, 509, 146

%\bibitem[\protect\citeauthoryear{Gonzalez Delgado \& Perez}{1997}]{GD97}
%Gonzalez Delgado R.M., Perez E., 1997, MNRAS, 284, 931

%\bibitem[\protect\citeauthoryear{Gonzalez Delgado et al.}{1997}]{gon97}
%Gonzalez Delgado R.M., Perez E., Tadhunter C., Vilchez J.M., 
%Rodriguez-Espinosa J.M., 1997, ApJS, 108, 155

%\bibitem[\protect\citeauthoryear{Greusard et al.}{2000}]{gre00}
%Greusard D., Friedli D., Wozniak H., Martinet L., Martin P., 2000,
%A\&AS, 145, 425

%\bibitem[\protect\citeauthoryear{Ho et al.}{1997}]{ho97}
%Ho L.C., Filippenko A.V., Sargent W.L.W., 1997, ApJ, 487, 591

%\bibitem[\protect\citeauthoryear{Ho \& Ulvestad}{2001}]{HU01}
%Ho L.C., Ulvestad J.S., 2001, ApJS, 133, 77

\bibitem[\protect\citeauthoryear{Huchra, Vogeley \&
Geller}{1999}]{huchra99}
Huchra J.P., Vogeley M.S., Geller M.J, 1999, ApJS, 121, 287

%\bibitem[\protect\citeauthoryear{Jungwiert, Combes \& Axon}{1997}]{JCA97}
%Jungwiert B., Combes F., Axon D.J., 1997, A\&AS, 125, 479

%\bibitem[\protect\citeauthoryear{Kollatschny \& Fricke}{1985}]{KF85}
%Kollatschny W., Fricke K.J., 1985, A\&A, 143, 393

%\bibitem[\protect\citeauthoryear{Kotilainen, Ward \& Williger}{1993}]{KWW93}
%Kotilainen J.K., Ward M.J., Williger G.M., 1993, MNRAS, 263, 655

%\bibitem[\protect\citeauthoryear{Kotilainen \& Ward}{1994}]{KW94}
%Kotilainen J.K., Ward M.J., 1994, MNRAS, 266, 953

\bibitem[\protect\citeauthoryear{Krips et al.}{2011}]{krips2011}
Krips, M., et al., 2011, ApJ, 736, 37

%\bibitem[\protect\citeauthoryear{Kurtz \& Mink}{1998}]{KM98}
%Kurtz M.J., Mink D.J., 1998, PASP, 110, 934

%\bibitem[\protect\citeauthoryear{Lawrence et al.}{1985}]{law85}
%Lawrence A., Watson M.G., Pounds K.A., Elvis M., 1985, MNRAS, 217, 685

%\bibitem[\protect\citeauthoryear{Maiolino, Risaliti \& Salvati}{1999}]{MRS99}
%Maiolino R., Risaliti G., Salvati M., 1999, A\&A, 341, L35

%\bibitem[\protect\citeauthoryear{Maiolino et al.}{2000}]{mai00}
%Maiolino R., Alonso-Herrero A., Anders S., Quillen A., Rieke M.J.,
%Rieke G.H., Tacconi-Garman L.E., 2000, ApJ, 531, 219

%\bibitem[\protect\citeauthoryear{Malkan, Gorjian \& Tam}{1998}]{MGT98}
%Malkan M.A., Gorjian V., Tam R., 1998, ApJS, 117, 25

%\bibitem[\protect\citeauthoryear{M{\' a}rquez et al.}{2003}]{mar03}
%M{\' a}rquez I., Masegosa J., Durret F., Gonz{\' a}lez Delgado R.M., Moles M.,
%Maza J., P{\' e}rez E., Roth M., 2003, A\&A, 409, 459

\bibitem[\protect\citeauthoryear{Mazzalay et al.}{2013}]{mazzalay13}
Mazzalay, X., Rodr{\'{i}}guez-Ardila, A., Komossa, S. \& McGregor, P. J. 2013, MNRAS, 430, 2411

\bibitem[\protect\citeauthoryear{McGregor et al.}{2003}]{mcgregor03}
McGregor, P. J. et al., SPIE, 2003, 4841, 1581, eds. Iye, M. \&
Moorwood, A. F. M.


%\bibitem[\protect\citeauthoryear{Meixner et al.}{1990}]{mei90}
%Meixner M., Puchalsky R., Blitz L., Wright M., Heckman T.,
%1990, ApJ, 354, 158

%\bibitem[\protect\citeauthoryear{Miyaji, Wilson \& Perez-Fournon}{1992}]{MWP92}
%Miyaji T., Wilson A.S., Perez-Fournon I., 1992, ApJ, 385, 137

\bibitem[\protect\citeauthoryear{Morganti et al.}{2005}]{morganti05}
Morganti R., Tadhunter C.N., \& Oosterloo T.A., 2005, A\&A, 444, L9

%\bibitem[\protect\citeauthoryear{Morris \& Ward}{1988}]{MW88}
%Morris S.L., Ward M.J., 1988, MNRAS, 230, 639

\bibitem[\protect\citeauthoryear{Mouri, Kawara and Taniguchi}{2000}]{mouri00}
Mouri H., Kawara K, \& Taniguchi Y., 2000, ApJ, 528, 186

%\bibitem[\protect\citeauthoryear{Mulchaey et al.}{1992}]{mul92}
%Mulchaey J.S., Tsvetanov Z., Wilson A.S., Perez-Fournon I., 1992,
%ApJ, 394, 91

%\bibitem[\protect\citeauthoryear{Mulchaey \& Regan}{1997}]{MR97}
%Mulchaey J.S., Regan M.W., 1997, ApJL, 482, L135

\bibitem[\protect\citeauthoryear{M\"uller S\'anchez et al.}{2009}]{muller09}
M\"uller S\'anchez, F., Davies, R.I., Genzel, R., et al.\ 2009, ApJ,
691, 749

\bibitem[\protect\citeauthoryear{M{\"u}ller-S{\'a}nchez et al.}{2011}]{muller11} 
M{\"u}ller-S{\'a}nchez, F., Prieto, M.A., Hicks, E.K.S., et
al.\ 2011, ApJ, 739, 69

%\bibitem[\protect\citeauthoryear{Mundell et al.}{1995a}]{mun95}
%Mundell C.G., Holloway A.J., Pedlar A., Meaburn J., Kukula M.J.,
%Axon D.J., 1995a, MNRAS, 275, 67

%\bibitem[\protect\citeauthoryear{Mundell et al.}{1995b}]{mun95b}
%Mundell C.G., Pedlar A., Axon D.J., Meaburn J., Unger S.W., 1995b,
%MNRAS, 277, 641

%\bibitem[\protect\citeauthoryear{Mundell et al.}{2003}]{mun03}
%Mundell C.G., Wrobel J.M., Pedlar A., Gallimore J.F., 2003, ApJ, 583, 192

%\bibitem[\protect\citeauthoryear{Nagar et al.}{1999}]{nag99}
%Nagar N.M., Wilson A.S., Mulchaey J.S., Gallimore J.F., 1999, ApJS, 120, 209

%\bibitem[\protect\citeauthoryear{Nelson \& Whittle}{1995}]{nw95}
%Nelson C.H. Whittle M., 1995, ApJS, 99, 67

%\bibitem[\protect\citeauthoryear{Nelson \& Whittle}{1996}]{nw96}
%Nelson C.H., Whittle M., 1996, ApJ, 465, 96

%\bibitem[\protect\citeauthoryear{Nelson et al.}{2000}]{nelson00}
%Nelson C.H., Weistrop D., Hutchings J.B., Crenshaw D.M., Gull T.R.,
%Kaiser M.E., Kraemer S.B., Lindler D., 2000, ApJ, 531, 257

%\bibitem[\protect\citeauthoryear{Nilson P.}{1973}]{nil73}
%Nilson P., 1973, Uppsala General Catalogue of Galaxies

%\bibitem[\protect\citeauthoryear{Oliva et al.}{1995}]{oli95}
%Oliva E., Origlia L., Kotilainen J.K., Moorwood A.F.M., 1995, A\&A, 301, 55

%\bibitem[\protect\citeauthoryear{Oliva et al.}{1999}]{oli99}
%Oliva E., Origlia L., Maiolino R., Moorwood A.F.M., 1999, A\&A, 350, 9

\bibitem[\protect\citeauthoryear{Osterbrock \& Ferland}{2006}]{osterbrock06}
Osterbrock, D.E., \& Ferland, G.J., 2006, Astrophysics of gaseous
nebulae and active galactic nuclei, 2nd.~ed.~by D.E.~Osterbrock and
G.J.~Ferland.~Sausalito, CA: University Science Books, 2006



%\bibitem[\protect\citeauthoryear{Peterson et al.}{2004}]{pet04}
%Peterson B.M., et al., 2004, ApJ, 613, 682

\bibitem[\protect\citeauthoryear{Pier et al.}{1994}]{pier1994}
Pier, E.A., Antonucci, R., Hurt, T., Kriss, G., \& Krolik, J.,
1994, ApJ, 428, 124

%\bibitem[\protect\citeauthoryear{Plummer}{1911}]{pl11} Plummer H.C.,
%1911, MNRAS, 71, 460

%\bibitem[\protect\citeauthoryear{Pogge}{1989}]{pog89}
%Pogge R.W., 1989, ApJS, 71, 433

\bibitem[\protect\citeauthoryear{Prieto et al.}{2010}]{prieto10}
Prieto, M. A., Reunanen, J., Tristram, K. R. W., et al., 2010, MNRAS, 402, 724

\bibitem[\protect\citeauthoryear{Proga et al.}{2000}]{proga00}
Proga, D., Stone, J.M., \& Kallman, T.R., 2000, ApJ, 543, 686

\bibitem[\protect\citeauthoryear{Proga \& Kallman}{2004}]{proga04}
Proga, D., \& Kallman, T.R., 2004, ApJ, 616, 688

%\bibitem[\protect\citeauthoryear{Quillen et al.}{1999}]{qui99}
%Quillen A.C., Alonso-Herrero A., Rieke M.J., Rieke G.H., Ruiz M.,
%Kulkarni V., 1999, ApJ, 527, 696

%\bibitem[\protect\citeauthoryear{Regan \& Mulchaey}{1999}]{RM99}
%Regan M.W. Mulchaey J.S., 1999, AJ, 117, 2676

%\bibitem[\protect\citeauthoryear{Riffel et al.}{2006}]{rif06}
%Riffel R.A., Storchi-Bergmann T., Winge C., Barbosa F.K.B., 2006,
%MNRAS, 373, 2

\bibitem[\protect\citeauthoryear{Riffel \& Storchi-Bergmann}{2011a}]{mrk1066}
Riffel, Rogemar A. \& Storchi-Bergmann, T., 2011, MNRAS, 411, 469

\bibitem[\protect\citeauthoryear{Riffel \& Storchi-Bergmann}{2011b}]{mrk1157}
Riffel, R.A. \& Storchi-Bergmann, T., 2011, MNRAS, 417, 2752

\bibitem[\protect\citeauthoryear{Riffel, Storchi-Bergmann \& Winge}{2013a}]{mrk79}
Riffel, R.A., Storchi-Bergmann, T., \& Winge, C., 2013, MNRAS, 430, 2249

\bibitem[\protect\citeauthoryear{Riffel et al.}{2013b}]{riffel13}
Riffel, Rogemar A., Storchi-Bergmann, Vale, T. B., McGregor, P.
2014, MNRAS, in press (Paper~I)

%\bibitem[\protect\citeauthoryear{Rodriguez Espinosa et al.}{1996}]{rod96}
%Rodriguez Espinosa J.M., Perez Garcia A.M., Lemke D., Meisenheimer K.,
%1996, A\&A, 315, L129

%\bibitem[\protect\citeauthoryear{Salamanca et al.}{1994}]{sal94}
%Salamanca I., et al., 1994, A\&A, 282, 742

%\bibitem[\protect\citeauthoryear{Salvati et al.}{1993}]{sal93}
%Salvati M., et al., 1993, A\&A, 274, 174

%\bibitem[\protect\citeauthoryear{Sandage}{1961}]{san61}
%Sandage 1961

%\bibitem[\protect\citeauthoryear{Schinnerer, Eckart \& Tacconi}{2000}]{SET00}
%Schinnerer E., Eckart A., Tacconi L.J., 2000, ApJ, 533, 826

%\bibitem[\protect\citeauthoryear{Schmitt \& Kinney}{1996}]{SK96}
%Schmitt H.R., Kinney A.L., 1996, ApJ, 463, 498

%\bibitem[\protect\citeauthoryear{Schmitt et al.}{1999}]{sch99}
%Schmitt H.R., Storchi-Bergmann T., Fernandes R.C., 1999, MNRAS, 303, 173

%\bibitem[\protect\citeauthoryear{Schmitt et al.}{2001}]{sch01}
%Schmitt H.R., Ulvestad J.S., Antonucci R.R.J., \& Kinney A.L., 2001,
%ApJS, 132, 199

%\bibitem[\protect\citeauthoryear{Schmitt et al.}{2003}]{sch03}
%Schmitt H.R., Donley J.L., Antonucci R.R.J., Hutchings J.B., \& Kinney A.L.,
%2003, ApJS, 148, 327

%\bibitem[\protect\citeauthoryear{Singh}{1999}]{sin99}
%Singh K.P., 1999, MNRAS, 309, 991

%\bibitem[\protect\citeauthoryear{Sosa-Brito et al.}{2001}]{sos01}
%Sosa-Brito R.M., Tacconi-Garman L.E., Lehnert M.D., Gallimore J.F.,
%2001, ApJS, 136, 61

%\bibitem[\protect\citeauthoryear{Storchi-Bergmann et al.}{2000}]{sto00} 
%Storchi-Bergmann T., Raimann D., Bica E.L.D., Fraquelli H.A., 2000,
%ApJ, 544, 747

%\bibitem[\protect\citeauthoryear{Storchi-Bergmann et al.}{2001}]{sto01} 
%Storchi-Bergmann T., Gonz{\' a}lez Delgado R.M., Schmitt H.R.,
%Cid Fernandes R., Heckman T., 2001, ApJ, 559, 147

\bibitem[\protect\citeauthoryear{Storchi-Bergmann et al.}{2009}]{sb09}
Storchi-Bergmann, T., McGregor, P. J.,  Riffel, R. A., Sim\~oes
Lopes, R., Beck, T., \& Dopita, M., 2009, MNRAS, 394, 1148

\bibitem[\protect\citeauthoryear{Storchi-Bergmann et al.}{2010}]{sb10}
Storchi-Bergmann, T., Sim\~oes Lopes, R., McGregor, P. Riffel,
Rogemar A., Beck, T., Martini, P., 2010, MNRAS, 402, 819

\bibitem[\protect\citeauthoryear{Storchi-Bergmann et al.}{2012}]{storchi12} 
Storchi-Bergmann T., Riffel R.A., Riffel R., et al., 2012, ApJ, 755, 87

%\bibitem[\protect\citeauthoryear{Terlevich et al.}{1990}]{tdt90}
%Terlevich E., Diaz A.I. Terlevich R., 1990, MNRAS, 242, 271

%\bibitem[\protect\citeauthoryear{Tonry \& Davis}{1979}]{TD79}
%Tonry J.L. Davis M., 1979, AJ, 84, 1511

%\bibitem[\protect\citeauthoryear{Tremaine et al.}{2002}]{tre02}
%Tremaine S., et al., 2002, ApJ, 574, 740

%\bibitem[\protect\citeauthoryear{Ulvestad \& Wilson}{1984}]{UW84}
%Ulvestad J.S., Wilson A.S., 1984, ApJ, 285, 439


%\bibitem[\protect\citeauthoryear{de Vaucouleurs et al.}{1991}]{deV91}
%de Vaucouleurs G., de Vaucouleurs A., Corwin H.G., Buta R.J.,
%Paturel G., Fouque P., 1991, Volume 1-3, XII, 2069 pp. 7
%figs..  Springer-Verlag Berlin Heidelberg New York

%\bibitem[\protect\citeauthoryear{Veilleux, Tully, \&
%Bland-Hawthorn}{1993}]{VTB93} Veilleux S., Tully R.B.,
%Bland-Hawthorn J., 1993, AJ, 105, 1318

%\bibitem[\protect\citeauthoryear{Veilleux}{1991}]{vei91}
%Veilleux S., 1991, ApJS, 75, 383

%\bibitem[\protect\citeauthoryear{Veilleux et al.}{1997}]{vei97}
%Veilleux S., Goodrich R.W., Hill G.J., 1997, ApJ, 477, 631

%\bibitem[\protect\citeauthoryear{Veilleux et al.}{2003}]{vei03}
%Veilleux S., Shopbell P.L., Rupke D.S., Bland-Hawthorn J., Cecil G.,
%2003, AJ, 126, 2185

\bibitem[\protect\citeauthoryear{Veilleux et al.}{2005}]{veilleux05}
Veilleux S., Cecil G., \& Bland-Hawthorn J., 2005, ARA\&A, 43, 769

%\bibitem[\protect\citeauthoryear{Weaver Gelbord \& Yaqoob}{2001}]{WGY01}
%Weaver K.A., Gelbord J., Yaqoob T., 2001, ApJ, 550, 261

%\bibitem[\protect\citeauthoryear{Winge et al.}{1995}]{win95}
%Winge C., Peterson B.M., Horne K., Pogge R.W., Pastoriza M.G.,
%Storchi-Bergmann T., 1995, ApJ, 445, 680

%\bibitem[\protect\citeauthoryear{Winkler et al.}{1992}]{win92}
%Winkler H., Glass I.S., van Wyk F., Marang F., Jones J.H.S.,
%Buckley D.A.H., Sekiguchi K., 1992, MNRAS, 257, 659

\bibitem[\protect\citeauthoryear{Woo \& Urry}{2002}]{woo02}
Woo, J.-H., \& Urry, C.~M., 2002, ApJ, 579, 530

%\bibitem[\protect\citeauthoryear{Wozniak et al.}{2003}]{woz03}
%Wozniak H., Combes F., Emsellem E., Friedli D., 2003, A\&A, 409, 469

%\bibitem[\protect\citeauthoryear{Young \& Devereux}{1991}]{YD91}
%Young J.S., Devereux N.A., 1991, ApJ, 373, 414

\end{thebibliography}
\end{document}